\documentclass{aa}  

\usepackage{graphicx}
\usepackage{txfonts}
\usepackage{natbib}
\usepackage{aas_macros}
\usepackage{enumitem}
\usepackage{booktabs} 
\newcommand{\ra}[1]{\renewcommand{\arraystretch}{#1}}
\usepackage[mathscr]{euscript} 

\usepackage{hyperref}

\begin{document} 


\title{Localising fast radio bursts and other transients using interferometric arrays}

\author{M. Obrocka\inst{1}
          \and
          B. Stappers\inst{1}
          \and 
	  P. Wilkinson\inst{1}}

\institute{Jodrell Bank Centre for Astrophysics, School of Physics and
		Astronomy, The University of Manchester, Manchester M13 9PL,UK \\
              \email{obrocka@jb.man.ac.uk}
             }

   \date{Received December, 2014; accepted February 2015}


\abstract{A new population of sources emitting fast and bright transient radio
bursts, called FRBs, has recently been identified. The observed large
dispersion measure values of FRBs suggests an extragalactic origin and an
accurate determination of their positions and distances will provide an unique
opportunity to study the magneto-ionic properties of the intergalactic medium.
So far, FRBs have all been found using large dishes equipped with multi-pixel
arrays. While large single dishes are well-suited for the discovery of
transient sources they are poor at providing accurate localisations. A
two-dimensional snapshot image of the sky, made with a correlation
interferometer array, can provide an accurate localisation of many compact
radio sources simultaneously. However, the required time resolution to detect
FRBs and a desire to detect them in real time, makes this currently
impractical. In a beamforming approach to interferometry, where many narrow
tied-array beams (TABs) are produced, the advantages of single dishes and
interferometers can be combined.  We present a proof-of-concept analysis of a
new non-imaging method that utilises the additional spectral and comparative
spatial information obtained from multiple overlapping TABs to estimate a
transient source location with up to arcsecond accuracy in almost real time. We
show that this method can work for a variety of interferometric configurations,
including for LOFAR and MeerKAT, and that the estimated angular position may be
sufficient to identify a host galaxy, or other related object, without
reference to other simultaneous or follow-up observations. With this method,
many transient sources can be localised to small fractions of a
half-power-beamwidth (HPBW) of a TAB, in the case of MeerKAT, sufficient to
localise a source to arcsecond accuracy. In cases where the position is less
accurately determined we can still significantly reduce the area that need be
searched for associated emission at other wavelengths and potential host
galaxies.}

\maketitle


\section{Introduction}
In the past few years, a possible new population of sources emitting fast and
bright transient radio bursts has been discovered. In August 2001 an isolated
pulse of radio emission was recorded but was only discovered during the
analysis of the archival survey data of the Magellanic clouds \citep{mfl+06}.
The discovery was published in 2007 \citep{lbm+07} and has since been called
the \emph{Lorimer Burst}. Another similar transient was found 5 years later
during reanalysis of the Parkes Multi-beam Pulsar Survey (PMPS) data
\citep{kskl12}. Four more, so-called \emph{fast radio bursts} (FRBs), were
found during analysis of the High Time Resolution Universe (HTRU) survey also
conducted with the Parkes telescope \citep{tsb+13}, which confirmed this class
of sources. Recently, a new FRB has been discovered in the 1.4-GHz Pulsar ALFA
survey conducted with the Arecibo Telescope \citep{sch+14}. It is the first FRB
observed at a different geographical location than the Parkes telescope, which
further confirms them as a class of sources and distinguishes them from the
perytons \citep{bsb11}. Recently three more FRBs were found
\citep{pbb+14,rsj14,bsb14}.\\ 


So far all detected FRBs have been found with single dish telescopes
and where they occurred within the beam of these telescopes is unconstrained.
Not only does this limit the ability to localise them it also limits our
ability to know their intrinsic luminosity, polarisation and spectral index
which in turn affects our understanding of their origin, emission mechanism and
their use as cosmological probes. So far the limited evidence available
suggests that the spectra of FRBs are flat \citep{tsb+13,hkf13}, but due to the
frequency-dependent gain pattern of a radio telescope\footnote{The
frequency-dependent primary beam corrections for a single dish are valid only
to the first null. The sidelobes beyond that are not characterised.}, a source
detected off-axis can have its spectral index distorted substantially. For
example, \cite{sch+14} reported a spectral index of $+7\pm1$ for FRB 121102
discovered with Arecibo. They created a map of the apparent instrumental
spectral index for the ALFA receiver using the gain variation in units of
$\text{K\;Jy}^{-1}$ and were able to show that an observed spectral index (if
not intrinsic) could only occur if the source was detected on the rising edge
of the first sidelobe.\\

The observed large dispersion measure (DM) values of the FRBs along the line of
sight implies large integrated electron density values that greatly exceed the
modelled predictions for the contribution from our Galaxy
\citep{col02}. This suggests an extragalactic origin for FRBs. Follow
up radio observations have failed to find further bursts at the same location
and it is therefore believed FRBs are one-off events that occasionally
"light-up" the radio sky. The estimated event rate is thousands per day
\citep{tsb+13,lkm+13}. As FRBs might represent a new class of highly compact
and extreme events in the Universe, it is vital to pin down the source location
to the arcsecond level in order to identify a potential host galaxy at high
redshift. This is especially true if there is no afterglow, or other associated
emission, at any other wavelength that might help to reveal the location
sufficiently precisely.\\ 

As well as identifying a new class of source, an exact determination of FRB
positions and distances (via the redshift of the presumed host galaxy) will
provide an unique opportunity to study the magneto-ionic properties of the
intergalactic medium (IGM) \citep{mk13}. It is thought that a substantial
fraction of the $60\%$ of the missing baryons in the Universe resides in
intergalactic gas \citep{co99}. The characteristics of the missing baryons are
difficult to constrain if they occur at densities and temperatures that do not
show significant absorption or emission. One possibility is if there
is an association between FRBs and Gamma-ray bursts (GRB) a new window on
cosmology will open. From such an association two precise measurements can be
made, namely the DM from the FRB and the redshift of the system from the GRB
\citep{dz14}. Together these will allow us to directly measure the IGM portion
of the baryon mass fraction, which in turn will constrain the re-ionisation of He
and H in the Universe. Other methods of obtaining the redshift, such as
associated multi-wavelength emission or host galaxy identification, would also
enable these measurements. Thus far, the association between FRBs
and GRBs has not been found despite dedicated searches
\citep{bmg+12,pwt+14}.\\ 

When larger FRBs samples (>100) are available with arcminute positional
precision at redshifts greater than 0.5, a stacking analysis technique might be
able to be used to study the baryonic mass profile surrounding different galaxy
types \citep{mcq14}. Even larger samples can be used to constrain the dark
energy equation of state \citep{glz14,zlw+14}.\\ 

None of these exciting opportunities can be grasped without much bigger and
better localised samples of FRBs than now exist. Localising while discovering
is key as the events are not repeatable in nature. A new observing strategy will
have to be developed to conduct surveys capable of locating large numbers of
FRBs. Describing such a strategy is the purpose of this paper.\\

\section{Approaches to the problem}
High time resolution radio astronomy has been, for a long time, dominated by
large filled-aperture dishes. The dish size dictates an area of sky seen in a
single pointing and while a large dish increases the sensitivity at the same
time it reduces the instantaneous field-of-view (FoV). To discover
large numbers of FRBs a wide FoV is desirable. Again, the accuracy of a source
location is approximately limited to an area covered by the HPBW and so
narrower beams are necessary for localisation. Even the biggest single dish, the
Five-hundred-meter Aperture Spherical radio Telescope (FAST) \citep{nlj+11},
currently being built in China, will achieve a positional accuracy of only 3.4
arcminute at 1.4 GHz.  This is insufficient to identify a transient in the
absence of further information.\\ 

To compensate for the small instantaneous FoV, a multi-pixel array of antennas
can be placed at the focal plane of a dish. The last two decades have seen
advances in the use of feed horn cluster receivers
\citep{mlc+01,cfl+06} which increase the survey speed while
maintaining the angular resolution dictated by the dish size. The independent
and static beams formed by the horns, while increasing the FoV, are spaced
further apart on the sky than the Nyquist theorem dictates for uniform
sampling. This can be improved with phased array feeds (e.g.
\cite{ovv10,cor+10}) of densely packed small antenna elements coupled with
beamforming networks that can synthesise multiple simultaneous and, if needed,
overlapping beams. It is therefore clear that large single dishes are
well-suited for the discovery of transient sources but are poor at localising
them.\\

To achieve even better sensitivity, to preserve the FoV and yet have high angular
resolution, the next generation radio telescopes are typically interferometers
made up of many relatively small elements, where the signals are combined
across the array to emulate a much larger dish. A two-dimensional snapshot image
of the sky, made with a correlation interferometer array, can provide a large
FoV and an accurate localisation of many compact radio sources simultaneously.
However since hundreds of measurements (baselines) may be acquired for the
image the data are often integrated over times of a few seconds; the
information about the sub-second variability in the sky is therefore lost. Fast
imaging methods that integrate the data on time scales of less than one second
have been shown \citep{ljb+11}, or more recently, \cite{lbb+14}
demonstrated an interferometric imaging campaign tailored to searching for FRBs
with 5 ms time resolution using the VLA. Still, the problem of data management
and the high computational load remain. We note that the computational and data
rates for different transient search methods are discussed in detail in papers
such as \cite{bc11} and \cite{lb14}. This limitation can be, to some extent,
overcome with a transient buffer a storage facility allowing only data with a
possible transient detection to be preserved for non-real-time analysis.
However, to explore the transient phase space, it is necessary to search out to
very high DMs, which places a demand on memory capacity to capture the fully
dispersed transient event. As the dispersion grows with $\nu^{-2}$ and the real
DM is unknown, the transient buffer capacity requirements can be large,
especially at low frequencies.  There is also a significant delay between the
detection and localisation for multi-wavelength identification of a transient
source when the data are imaged.  In short, the classical correlation approach
to interferometry is not well suited to finding FRBs.\\

For the joint task of discovery and simultaneous localisation, the advantages of
single dishes and interferometers are combined in a \emph{beamforming}
approach, where the input signals from antennas are added coherently producing
one or more narrow \emph{tied-array beams} (TABs). This coherent
addition of signals maximises instantaneous sensitivity that scales linearly
with the number of dishes in the array.
%
%
The relative localisation capabilities  of a \textit{single} TAB are similar to
a single dish, except it is narrower by a factor $B/d$, where \textit{B} is the
maximum baseline and \textit{d} is the size of the primary element,
where primary element is the smallest building block of the telescope;
e.g. dish or tile or antenna. The downside is that to achieve a similar FoV to
that of a correlation interferometer i.e. that of a primary element, a very
large number of TABs may have to be formed. The real-time signal processing of
a beamformer can therefore become computationally expensive. The sky coverage
may therefore have to be sacrificed and only the elements contained in an inner
core region of the array used. This leads to wider TABs with reduced
sensitivity.  Optimising this combination, including the total of the number of
TABs one can form and subsequently process, is an important consideration of
time domain radio astronomy using interferometers. But without imaging and with
only a single observation, a transient source position can still only be
approximated to within the HPBW of a TAB. In a case of a strong source, it
might even be detected in a sidelobe \citep{sch+14}. To overcome this
limitation we present a new method that utilises the additional spectral and
comparative spatial information obtained from multiple TABs to estimate a
transient source location with up to arcsecond accuracy in almost real time
without imaging.\\ 

\begin{figure*}
        \includegraphics[width=.9\linewidth/2]{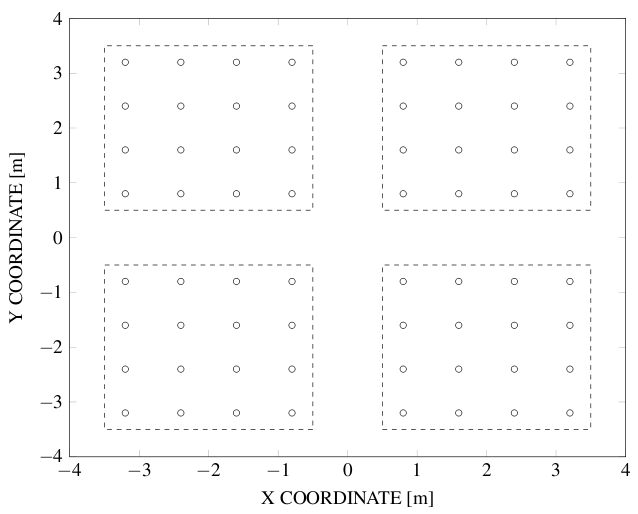}
	\includegraphics[width=.9\linewidth/2]{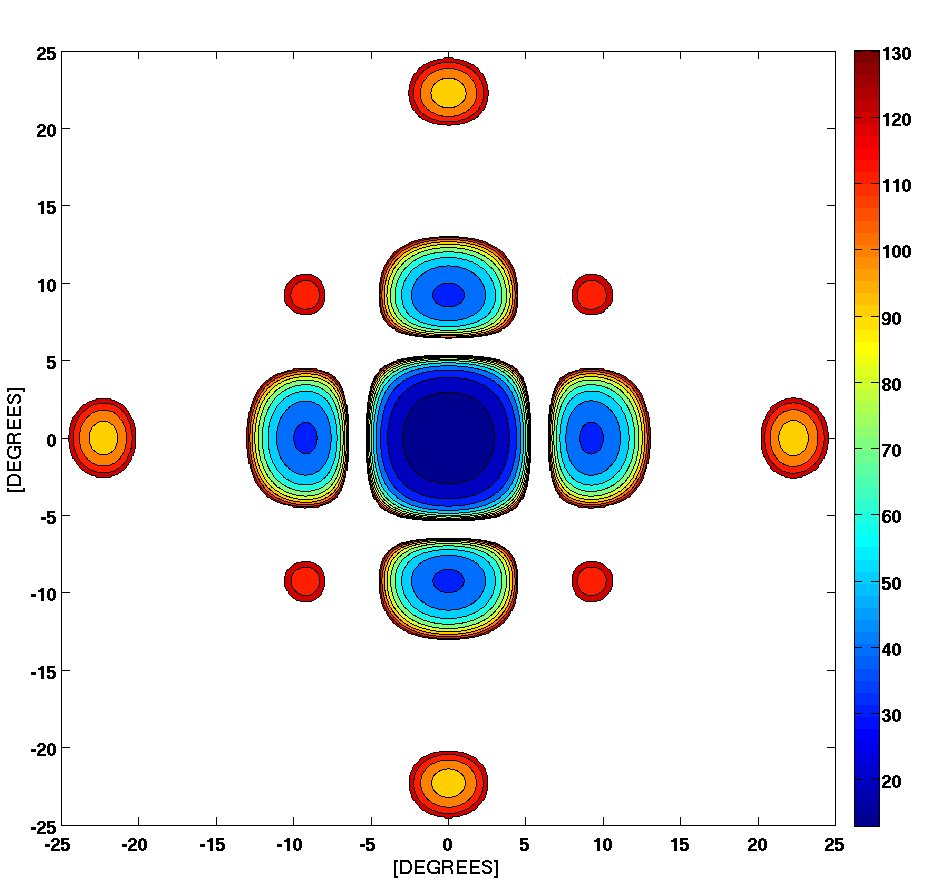}        
	
	\includegraphics[width=.9\linewidth/2]{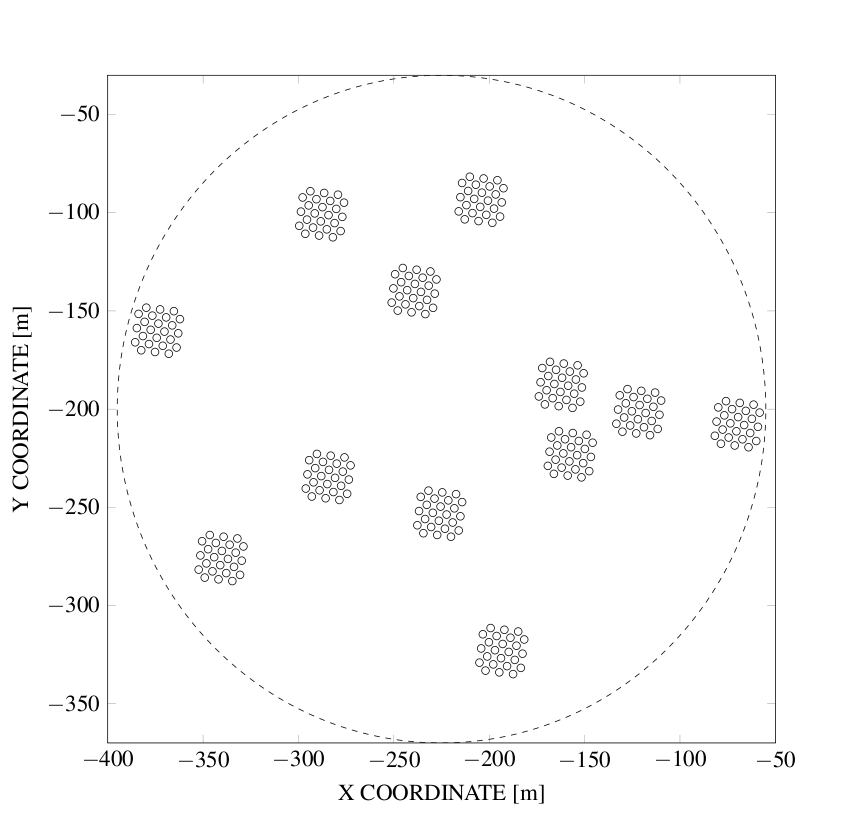}
        \includegraphics[width=.9\linewidth/2]{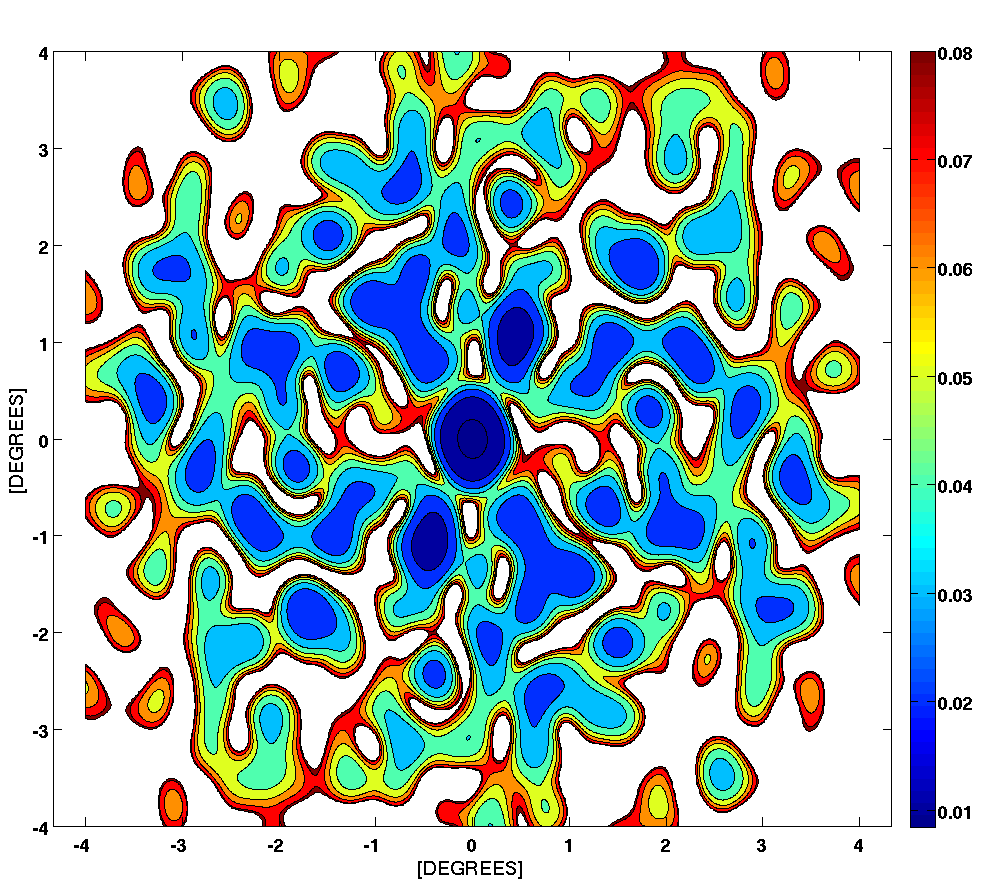}
	
	\includegraphics[width=.9\linewidth/2]{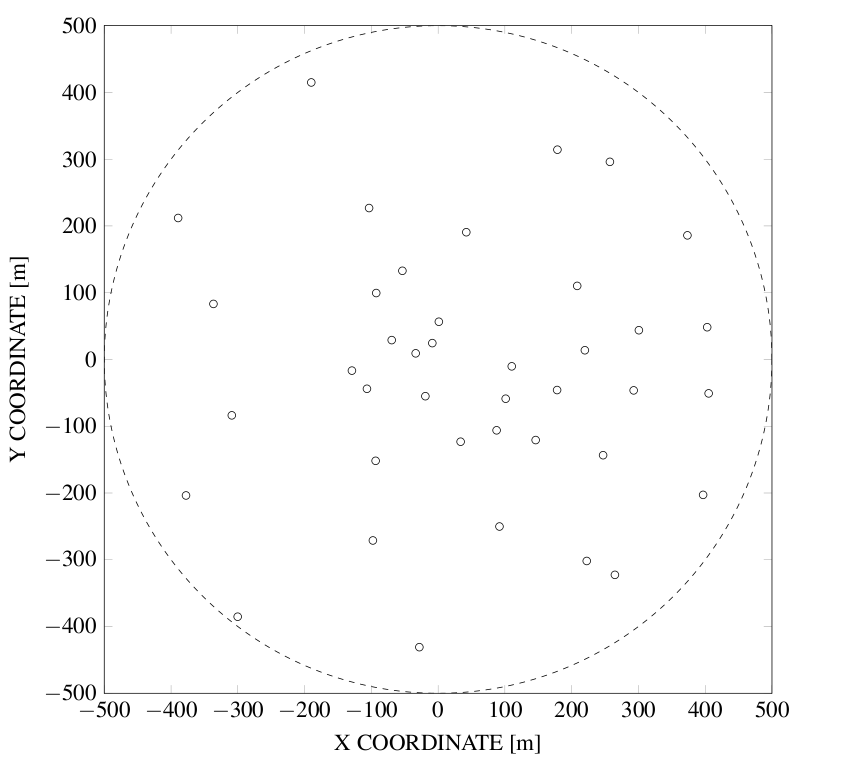}
	\includegraphics[width=.9\linewidth/2]{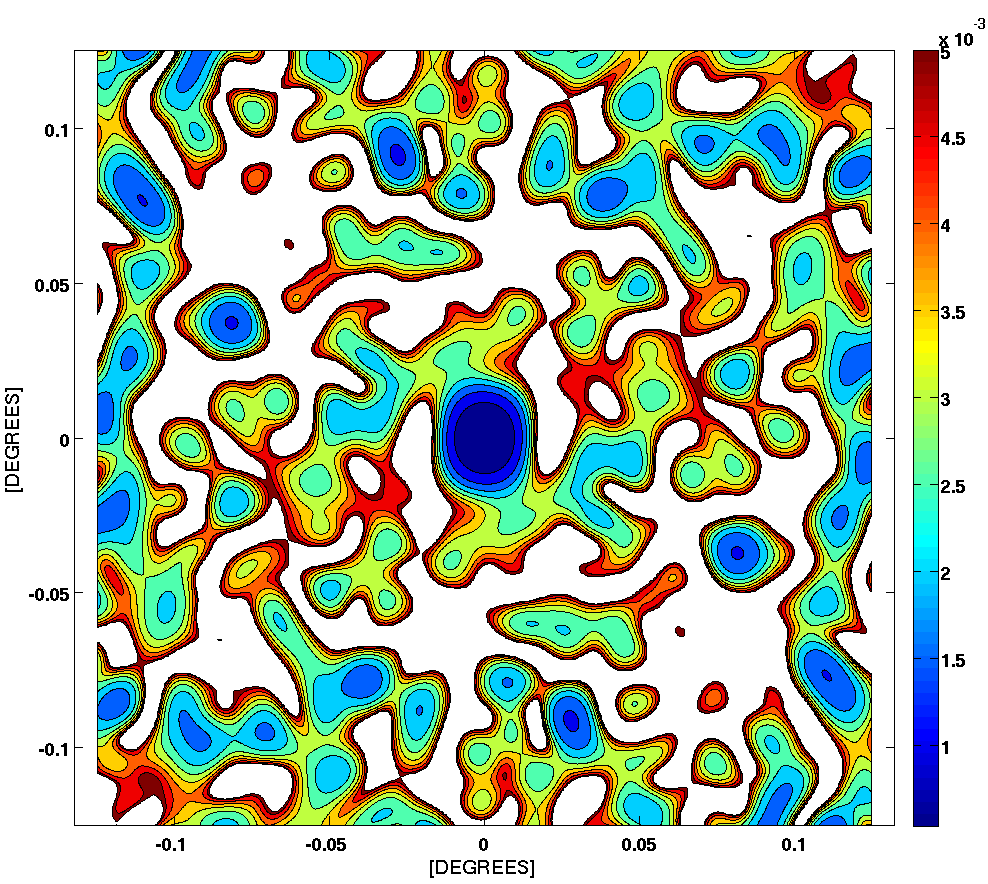}
\caption{Left column: array configuration and right column sensitivity
$\mathscr{S}$ maps, in units of Jy of the MUST, LOFAR and MeerKAT array
elements considered here. Top row: the MUST array at 575 MHz; middle row: the
LOFAR Superterp at 119 MHz; bottom row: the MeerKAT array core at 1400 MHz. In
all cases the antennas are assumed to be pointing at the zenith.}
\label{fig:sens_maps}
\end{figure*}

We consider, as examples for simulation, three different parts of new and
upcoming telescopes with multibeaming capabilities; (i) the MUST
array\footnote{The Manchester University Student Telescope; a small test array
at the Jodrell Bank observatory site that consists of four individual tiles
with Yagi antennas positioned on a regular $4\times4$ grid with 0.8 meter
spacing in each polarisation. The maximum separation between the tiles is 3.5
meters (centre to centre).}; (ii) the LOFAR Superterp \citep{vwg+13} and (iii)
the MeerKAT array core \citep{bdjf09}. The configurations of each array are
shown in Fig. \ref{fig:sens_maps} (left column). Only the inner regions of
MeerKAT and LOFAR are considered as they have a suitable sensitivity and FoV
combination. Each array operates at a different frequency, has different
baseline lengths, numbers of elements and types of antennas. The
comparison of the simulated results demonstrates the flexibility of our method
and illuminates its capabilities.\\

This paper is organised as follows. In Section 3 we describe the method for
utilising one-dimensional beamformed data to accurately estimate a transient
source location. In Section 4, we present the simulation results for the MUST,
MeerKAT and LOFAR arrays. Finally in Section 5 we discuss the results and in
Section 6 we present our conclusions.

\section{Method}
\begin{figure}
\includegraphics[width=\linewidth]{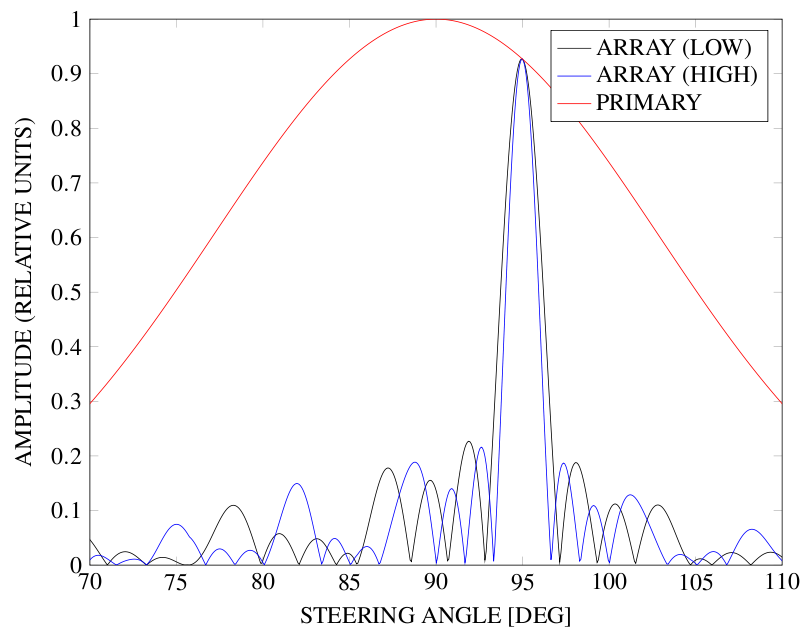}
\caption{Illustration of the frequency-dependent gain pattern variations for an
arbitrary array. The graph depicts a TAB at the lowest frequency (black) and at
the highest operational frequency (blue). The fractional bandwidth is $26\%$.
The primary beam (red) is plotted only at the lowest frequency for clarity.}
\label{fig:arb_steering}
\end{figure}

The independent TABs of an interferometer can be electronically steered to any
direction within the beam of the primary element. They can be arranged into any
pattern in the FoV, including Nyquist sampling, in contrast to the fixed beam
patterns from horn receivers on single dishes. The TAB gain pattern is, of
course, frequency dependent i.e. the beam width gets narrower and sidelobes
move closer to the main beam with increasing frequency, as illustrated in
Fig. \ref{fig:arb_steering}. These frequency-dependent variations in a TAB
shape can be used to create a sensitivity $\mathscr{S}$ map (a 2-D array in RA
and declination) at different frequencies, in particular at the upper $\nu_H$
and lower $\nu_L$ frequencies in the observing band. To create the sensitivity
$\mathscr{S}$ map for each array, the normalised beam patterns, simulated with
MATLAB\footnote{\url{http://www.mathworks.co.uk}} and the more efficient OSKAR-2
package\footnote{\url{http://www.oerc.ox.ac.uk/~ska/oskar2/}}, are scaled to
the minimum flux sensitivity $S_{\nu\text{,min}}$ at the phase centre of each
TAB. To calculate the minimum flux sensitivity $S_{\nu\text{,min}}$ we used the
modified radiometer equation adapted from \cite{lfl06}:
\begin{equation}
  \label{eq:srms}
        S_{\nu\text{,min}} = \beta\frac{T\;
        \left(S/N\right)_{min}}{G\sqrt{n_p\tau\Delta\nu}}
        \:[\text{mJy}]
\end{equation}
where G is the effective telescope gain ($\text{K}/\text{Jy}$), $n_p$ is the
number of polarisations summed, $\Delta\nu$ is the observing bandwidth (MHz),
$\tau$ is the integration time (s), \textit{T} is the system temperature (K),
$\left(\text{S}/\text{N}\right)_{min}$ is the minimum signal-to-noise ratio and
$\beta$ accounts for digitisation losses and were taken from \cite{kv01}. The
values used are listed in Table \ref{table:sensitivity}.\\ 

Sky temperature values vary as a function of Galactic latitude and longitude.
To account for that we used the \cite{hssw82} 408-MHz all-sky survey. The sky
temperature is scaled to the frequencies listed in Table
\ref{table:sensitivity} under $\Delta\nu$. The scaling assumes a temperature
spectral index of synchrotron radiation of $\alpha$ of -2.6 \citep{rr88b}. The
resultant two-dimensional sensitivity maps are plotted in Fig.
\ref{fig:sens_maps} (right column). We note that we present
sensitivities here only as a basis for calculations and that the method
actually considers predicted S/Ns and so are simply scaled by any differences
in values of the true array from those in Table \ref{table:sensitivity}.\\

\begin{table*}\centering
\caption{Values used in Equation \ref{eq:srms}, where $\beta$ accounts for
digitisation losses; $A_{eff}$ is the effective area; $\tau$ is the integration
time; $\Delta\nu$ is the observing bandwidth; $n_p$ is the number of
polarisations summed; $T$ the system temperature given only for the zenith,
where the sky temperature $T_{\text{sky}}$ is included in $T$. In addition we
list the maximum baseline $B$ and the resulting HPBW at the lowest frequency in
the band.}
\label{table:sensitivity}
\centering
\begin{tabular}{l c c c c c c c c}
\hline\hline
Array & $\beta$ & $A_{eff}$ & $\tau$ & $\Delta\nu$ & $n_p$ & $T$ & $B$ & HPBW$_L$\\
& & [$m^2$] & [s] & [MHz] &  & [K] & [m] &\\
\hline
MUST & 0.99& 36 & $5\times 10^{-3}$ & 575-625 & 2 & 207 & 5 & $5.39^{\circ}$\\
LOFAR & 0.66 & 8064 & $5\times 10^{-3}$ & 119-150 & 2 & 907 & 300 & $0.43^{\circ}$\\
MeerKAT & 0.66 & $\sim5000$& $5\times 10^{-3}$ & 1400-1700 & 2 & 30 & 1000 & $0.52'$\\
\hline
\end{tabular}
\end{table*}

The location of a transient source detected with any single-beam radio
dish can be constrained only to an area defined by the beam
pattern of that dish. In the case of a beam forming interferometer, the
source position is most likely to be within the area defined by a HPBW
contour of a TAB.  This is only a first order assumption as in reality
a strong source can be detected anywhere in the beam, including sidelobes. When
a source is detected in multiple TABs the resulting detection pattern can be
used to approximate the source position to an area smaller than the area of an
individual TAB. We will now go on to show that we can use this detection
pattern to further constrain the source position even if it was detected in a
sidelobe.\\

\begin{figure}
\includegraphics[width=\linewidth]{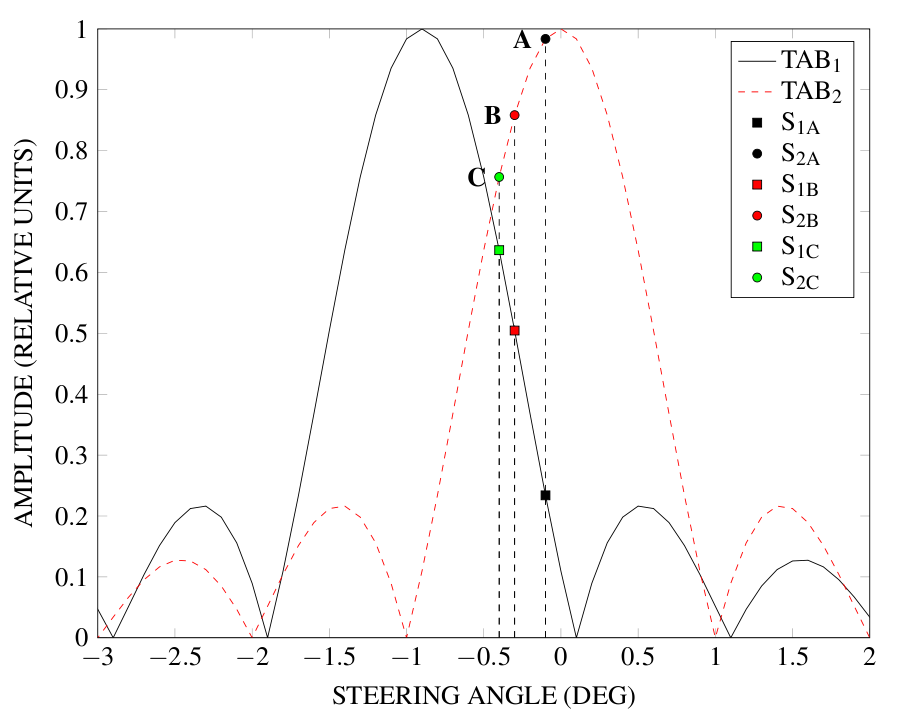}
\caption{Flux ratios in two overlapping TABs.
Three hypothetical strong sources are detected, their positions are
marked as A (black), B (red) and C (green). Note that the primary beam shape
is not included for clarity. The amplitudes have been normalised to peak at 1.}
\label{fig:mult_cap}
\end{figure}

\begin{figure}
\centering
	\includegraphics[width=\linewidth]{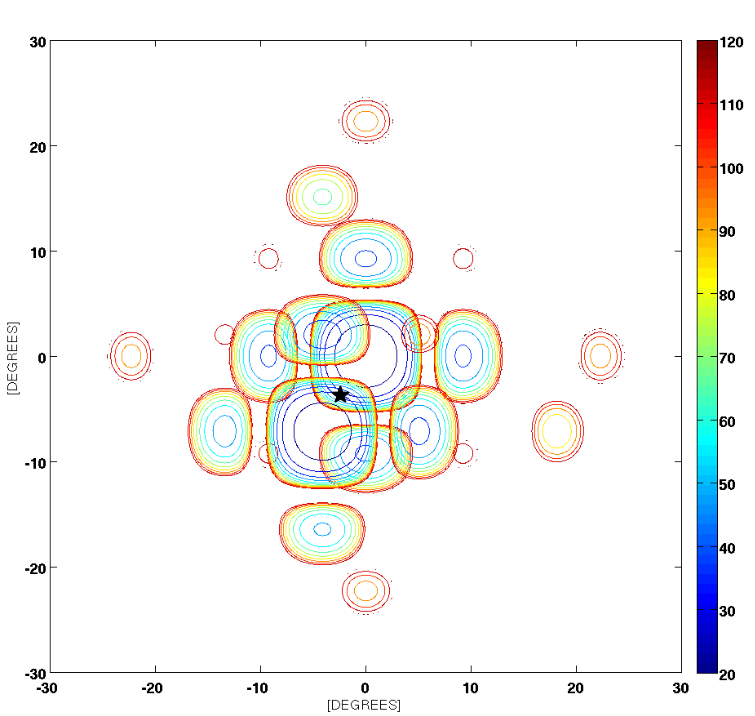}
	\includegraphics[width=\linewidth]{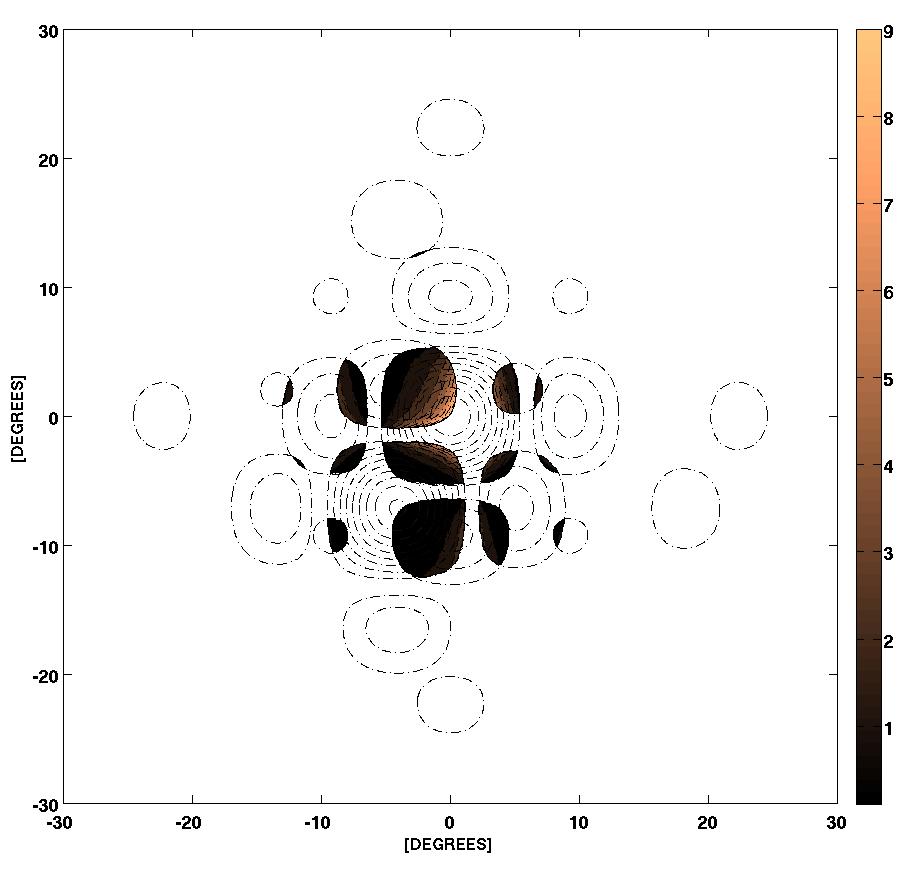}
\caption{(Top) contour plot of the sensitivity $\mathscr{S}$ maps of two
overlapping TABs and their sidelobes of the MUST array in units of Jy. The
black star depicts location of a source for a simple example in
\S\ref{sec:simple_example}. (Bottom) the flux density ratio
$\mathscr{S}_1/\mathscr{S}_2$ map with TAB contours plotted with dashed lines.
The flux ratio values are limited to the range $[0.1,10]$ for clarity.}
\label{fig:ratio_maps_must}
\end{figure}

The idea is that a pattern of the S/N  of the detection of a transient source
across the multiple TABs can be generated. From that pattern, corresponding
values of the observed (apparent) flux density $S_{\nu}$ can be calculated via
Equation \ref{eq:srms}. The value of the S/N and thus the observed $S_{\nu}$
depends on the source location within a TAB, as the sensitivity decreases away
from its phase centre. Fig. \ref{fig:mult_cap} illustrates a hypothetical
scenario, when three sources are detected in two overlapping TABs at positions
marked with A, B and C.  Each position is located further
away from the phase centre of each TAB. For $\text{TAB}_1$, the
location \textit{C} is the most sensitive position and for $\text{TAB}_2$
location \textit{A} is the most sensitive. The resulting ratios of the
normalised fluxes from both TABs are then:
\begin{equation*} 
S_{1A}/S_{2A} = 0.24 \;\text{,}\;S_{1B}/S_{2B} = 0.56 \;\; \text{and} \;\;
S_{1C}/S_{2C} = 0.84,
\end{equation*}
where we have assumed that all fluxes are above the detection threshold. Due to
the rapidly changing sensitivity of the TABs away from their phase centres, the
flux ratios at each position differ substantially. The ratio of the
\emph{observed} fluxes in the two TABs, $S_{1}/S_{2}$, can be compared with the
ratio of the \emph{calculated} sensitivity across each of TABs,
$\mathfrak{s}_{1}/\mathfrak{s}_{2}$, from the beam model and respective
sensitivity maps $\mathscr{S}_1$ and $\mathscr{S}_2$. By identifying the
locations where the observed and calculated flux ratios are the same, the
source location can be estimated with much higher precision than from a TAB on
its own. An example of a \emph{flux density ratio}
$\mathscr{S}_1/\mathscr{S}_2$ map is shown in Fig. \ref{fig:ratio_maps_must}
(bottom) for the simple MUST array and the glossary of symbols used is listed
in Table \ref{table:symbols}.\\ 

The problem is that the observed $S_{1}/S_{2}$ ratio may not be sufficient to
constrain the source location since the calculated
$\mathfrak{s}_{1}/\mathfrak{s}_{2}$ ratio can repeat in different areas of the
FoV, especially when the inevitable uncertainties are taken into account. Thus
another metric is needed.\\

\begin{table}
\caption{Glossary of symbols used in this paper.}
\label{table:symbols}
\centering
\begin{tabular}{l c c}
\hline\hline
Parameter & Flux density & Spectral index \\
\hline
   Observed (1-D)& \textit{S} & $\alpha$\\
   Map (2-D)& $\mathscr{S}$ & $\mathscr{A}$\\
   Map values (1-D)& $\mathfrak{s}$ & $\mathfrak{a}$\\
\hline
\end{tabular}
\end{table}


\begin{figure}
\centering
	\includegraphics[width=\linewidth]{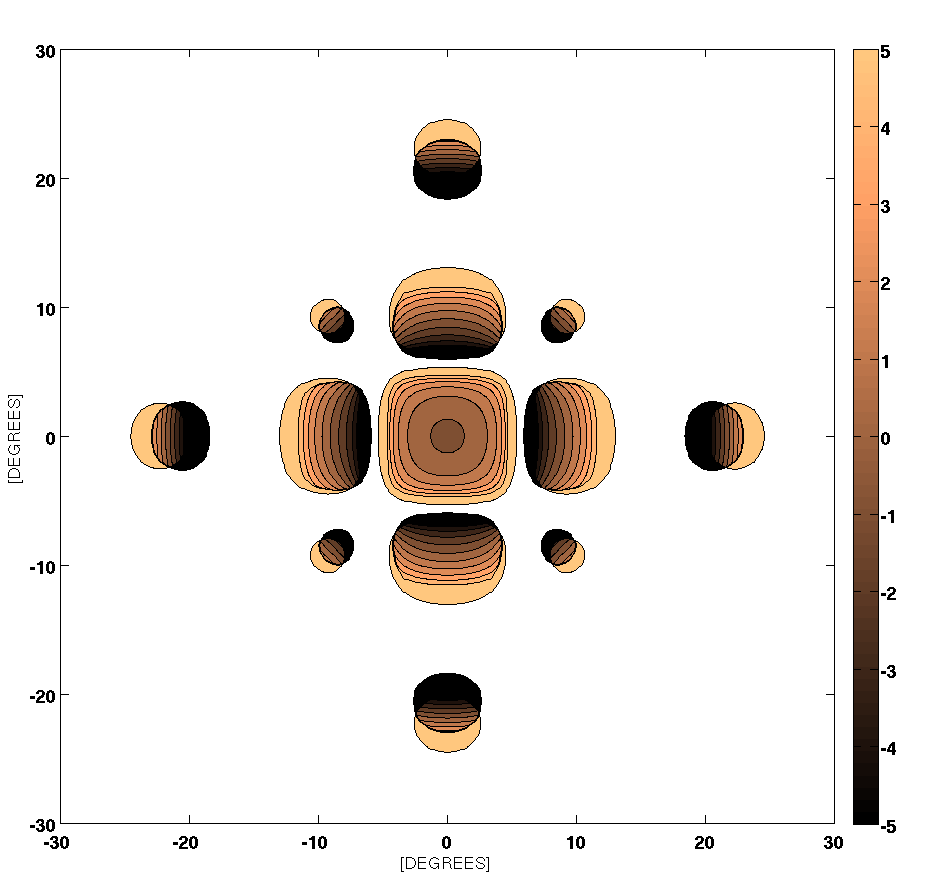}
	\includegraphics[width=\linewidth]{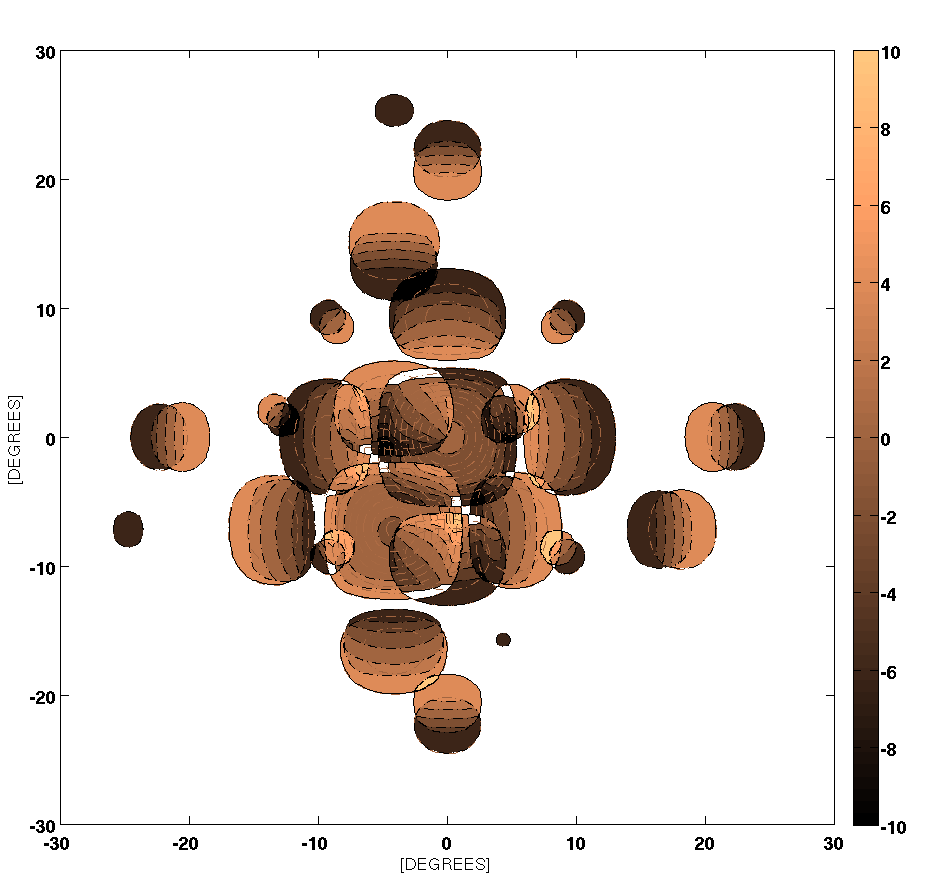}
\caption{The instrumental spectral index map $\mathscr{A}$ for the MUST array
(Top) the values of spectral index are limited to the range $(-5,5)$ and only
pixels with sensitivity $10\times S_{\nu\text{,min}}$ are displayed for
clarity. (Bottom) the difference spectral index $(\mathscr{A}_2 -
\mathscr{A}_1)$ map for the MUST array created from the TABs in Fig.
\ref{fig:ratio_maps_must}.}
\label{fig:must_array_patterns}
\end{figure}

The observed flux density $S_{\nu}$ from many radio sources follows a power law
dependency to the first order\footnote{At this stage of the analysis we are
considering only power laws.}:
\begin{equation}
\label{eq:si_f_ralationship}   
        S(\nu) \propto \nu^{\alpha},
\end{equation}
where $\alpha$ is the spectral index and $\nu$ is the observing frequency. If
observations at two distinct frequencies $\nu_L$ and $\nu_H$ are available, the
spectral index $\alpha$ can be calculated using Equation
\ref{eq:si_f_ralationship} as:
\begin{equation}
\label{eq:spectral_index_mod}   
        \alpha =
\frac{\log\left({\frac{S_L}{S_H}}\right)}{\log\left({\frac{\nu_L}{\nu_H}}\right)},
\end{equation}
where $S_L$ and $S_H$ are the apparent flux densities from observations at
frequencies $\nu_L$ and $\nu_H$. We note that when calculating these
sensitivities, we have assumed the full bandwidth given in Table
\ref{table:sensitivity} for each of $\nu_L$ and $\nu_H$. That is, $\nu_L$ and
$\nu_H$ are the centres of these bands. As discussed in Section 5 a subsequent
analysis should include the fact that $S_L$ and $S_H$ represent average values.
As above we note that the accuracy of our values for $S_L$ and $S_H$ do not
affect the efficacy of our method. Due to the frequency-dependent gain pattern
of a TAB, sources detected off-axis can have their spectral indices distorted
substantially. In our analysis we have assumed that the beam dependent spectral
index is also described by a power law. To investigate this effect further, we
use Equation \ref{eq:spectral_index_mod} and 2-D sensitivity $\mathscr{S}$ maps
for frequencies $\nu_L$ and $\nu_H$, to create a 2-D \emph{instrumental
spectral index}, $\mathscr{A}$, map of a telescope beam pattern, as illustrated
in Fig. \ref{fig:must_array_patterns} (top) for the MUST array.\\

In contrast to \cite{sch+14}, the instrumental spectral index $\mathscr{A}$ map
is created using the sensitivity (Equation \ref{eq:srms}) of the beam at a
given observing frequency rather than its gain. This is in order to include the
noise contribution from the sky. The apparent variation of the instrumental
spectral index away from the phase centre therefore has the opposite sign to
that of \cite{sch+14}. We assume that the observed spectral index $\alpha_{O}$
of a detected source is a combination of its intrinsic spectral index
$\alpha_{I}$ and the calculated instrumental spectral index $\mathfrak{a}$
imposed by the beam patterns. It can thus be described with the simple
equation:
\begin{equation}
\label{eq:si4}   
        \alpha_{I} = \alpha_{O} + \mathfrak{a}
\end{equation}
Using Fig. \ref{fig:mult_cap} as an illustration again, a source observed at
position A, with normalised flux density $S_{1A}$ from $\text{TAB}_1$ and
$S_{2A}$ from $\text{TAB}_2$, is also detected at both frequencies, $\nu_L$ and
$\nu_H$, yielding four different values of flux density. $S_{1A,\nu_L}$ and
$S_{1A,\nu_H}$ for $\text{TAB}_1$ and $S_{2A,\nu_L}$ and $S_{2A,\nu_H}$ for
$\text{TAB}_2$. Using Equation \ref{eq:spectral_index_mod} we can calculate the
observed spectral index $\alpha_{O_1}$ for source A detected in $\text{TAB}_1$
and $\alpha_{O_2}$ for source A detected in $\text{TAB}_2$. As both TABs
detected the same source A, Equation \ref{eq:si4} can be written as follows
(omitting subscript A for clarity):
\begin{align}
\label{eq:si1}
        \alpha_{O_1} = \alpha_{I} - \mathfrak{a}_1,\nonumber\\
        \alpha_{O_2} = \alpha_{I} - \mathfrak{a}_2.
\end{align}
These two equalities yield the relationship:
\begin{equation}
\label{eq:si3}   
        \alpha_{O_1} - \alpha_{O_2} = \mathfrak{a}_2 - \mathfrak{a}_1,
\end{equation} 
and the left side of this equality is known from the detections. To connect the
calculated difference $(\mathfrak{a}_2 - \mathfrak{a}_1)$ to a position within a
TAB we need to subtract the instrumental spectral index $\mathscr{A}_2$ map of
$\text{TAB}_2$ from the instrumental spectral index $\mathscr{A}_1$ map of
$\text{TAB}_1$. An example of such manipulation is illustrated in Fig.
\ref{fig:must_array_patterns} (bottom) for the MUST array. At any point the
difference of the observed spectral indices $(\alpha_{O_1} - \alpha_{O_2})$
together with the observed flux density ratio $S_{1}/S_{2}$, can be used to
constrain the position of a source within the two TABs with an accuracy
significantly better than a beam width. The symbols used in relation to the
spectral index manipulations are also summarised in Table \ref{table:symbols}.\\ 

\begin{figure}
        \includegraphics[width=\linewidth]{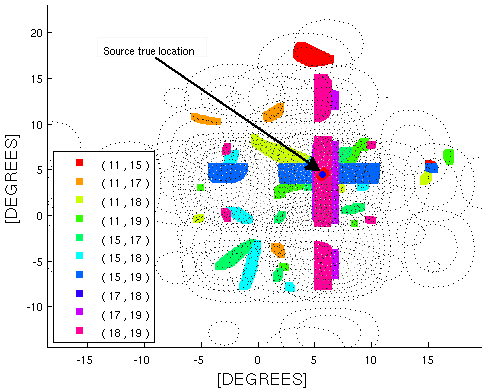}
\caption{Illustration of a detection made with five TABs that Nyquist
sample the FoV resulting in $C_2^5 = 10$ pair combinations. Each colour
depicts overlapping regions of $(\mathfrak{s}_1/\mathfrak{s}_2)_{(i,\;j)}$ and
$(\mathfrak{a}_2 - \mathfrak{a}_1)_{(i,\;j)}$ from each $(i,\;j)$ TAB pair.}
\label{fig:green_areas}
\end{figure}

The above two-TAB detection analysis of the observed flux density ratio and
spectral index difference produces a single or a set of possible locations for
a source. If a transient source was detected in \textit{N} TABs, this process
can be repeated for \emph{all $N(N-1)/2$ pair combinations}. To identify the
pair combinations we will use subscripts $(i,\;j)$. The final step in the
process is to compare the estimated locations resulting from all the TAB pairs
against each other. \emph{Only locations common for all pairs are chosen for
the estimated true  source position}. This is illustrated in Fig.
\ref{fig:green_areas}, for a source detected in five TABs that Nyquist
sample the FoV, where each colour depicts overlapping regions of
$(\mathfrak{s}_1/\mathfrak{s}_2)_{(i,\;j)}$ and $(\mathfrak{a}_2 -
\mathfrak{a}_1)_{(i,\;j)}$ from each TAB pair. For example, orange represents
overlapping regions of $(\mathfrak{s}_1/\mathfrak{s}_2)_{(11,\;17)}$ and
$(\mathfrak{a}_2 - \mathfrak{a}_1)_{(11,\;17)}$ for the $\text{TAB}_{11}$ and
$\text{TAB}_{17}$ pair. Crucially, only regions where all "colours" overlap are
treated as a possible source location. A simplified example is also given in
\S\ref{sec:simple_example}.\\

To summarise, our analysis we only consider sources that
were detected:
\begin{enumerate}[label=\emph{\alph*}),noitemsep,nolistsep]
   \item in at least two TABs and
   \item at two frequencies, $\nu_L$ and $\nu_H$ in each TAB.
\end{enumerate}
In reality a source could be detected in just one beam or at one frequency
only. Hence, our definition of detection is quite restrictive but is required
to provide better positional accuracy. A by-product of an accurate position
estimation is the possibility of recovering the intrinsic spectral index
$\alpha_I$ of a source. Since our main assumption is that a telescope imparts
deterministic instrumental spectral index to a detected source, the latter can
be corrected for if the position of the source within a TAB is known.\\

\begin{figure}
\centering
	\includegraphics[width=\linewidth]{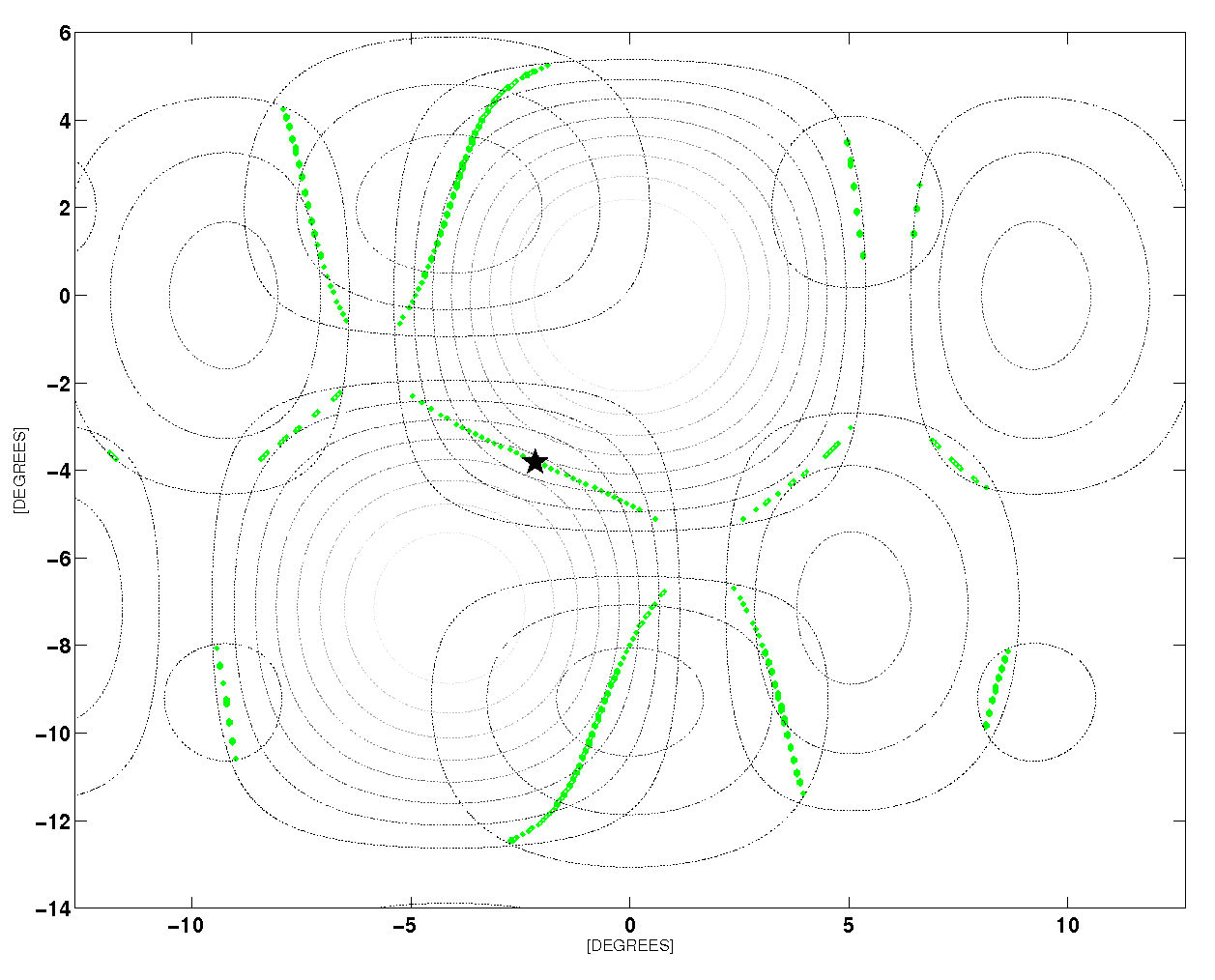}
	\includegraphics[width=\linewidth]{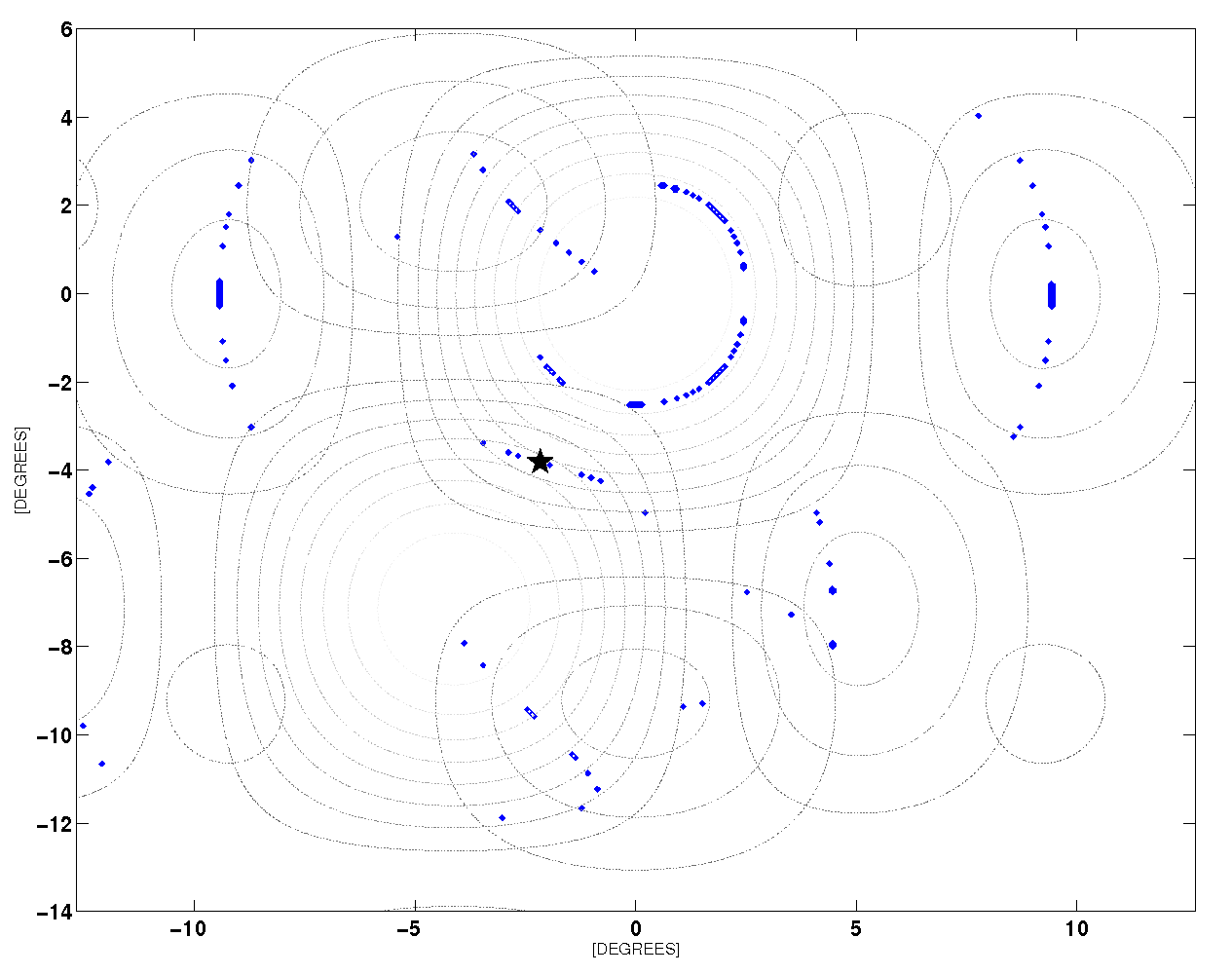}
        \includegraphics[width=\linewidth]{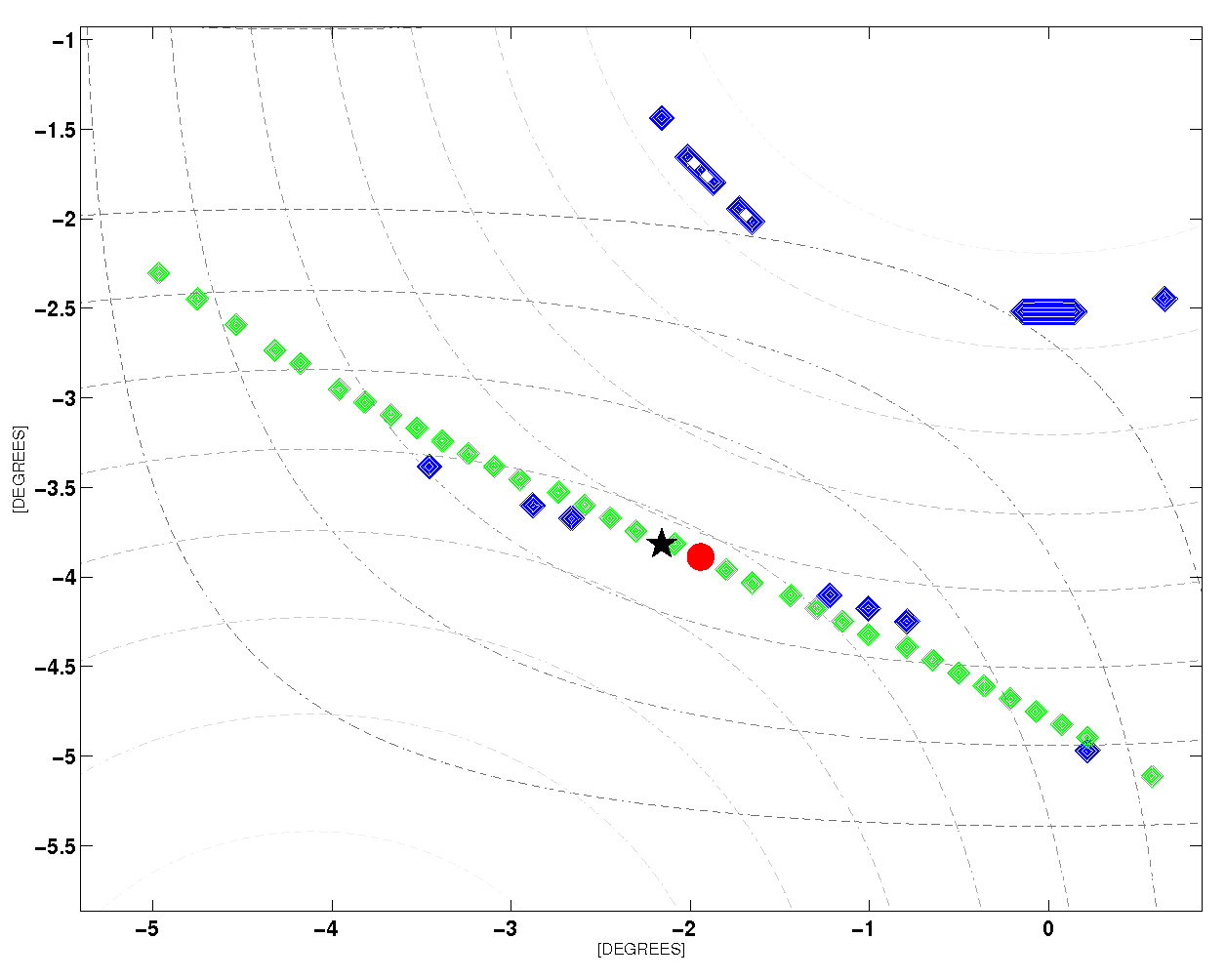}
\caption{A hypothetical source was detected in two overlapping MUST TABs at a
position $[-2.1^{\circ},-3.8^{\circ}]$, depicted as a black star. The top
figure shows the values of $\mathfrak{s}_1/\mathfrak{s}_2 = S_1/S_2 \pm1\%$,
plotted on top of the TAB contours. The middle figure shows values of
$(\mathfrak{a}_2 - \mathfrak{a}_1) = (\alpha_{O_1} - \alpha_{O_2}) \pm1\%$.
The bottom figure shows both $(\mathfrak{a}_2 - \mathfrak{a}_1)$ and
$\mathfrak{s}_1/\mathfrak{s}_2$ overlaid. The red dot at a position
$[-1.95^{\circ},-3.9^{\circ}]$ indicates the best estimated position.}
\label{fig:must_ratio_maps}
\end{figure}

\begin{figure}
\centering
\includegraphics[width=\linewidth]{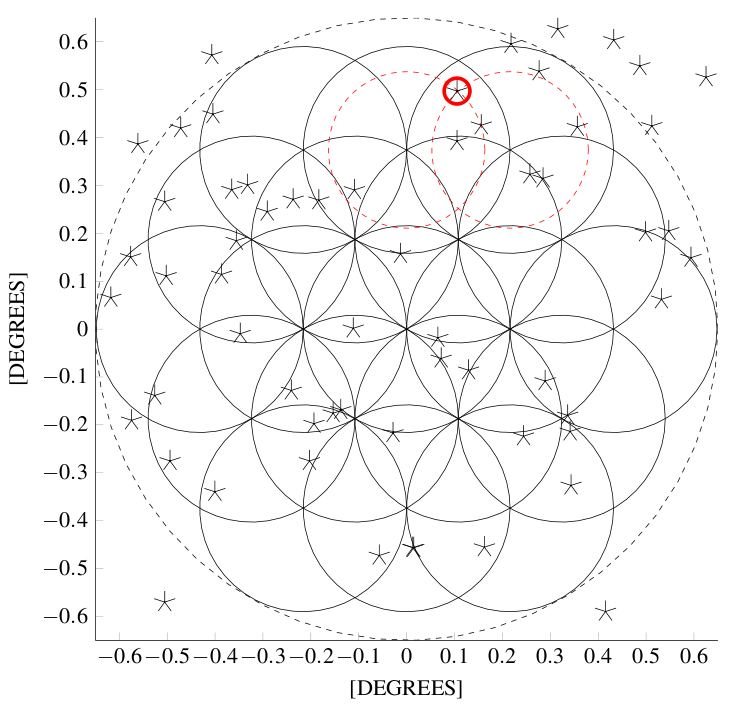}
\caption{A schematic of Nyquist sampled LOFAR TABs which illustrates the
general layout assumed in all simulations. Black circles represent a 3-dB
contour of a TAB at the lower frequency $\nu_L$ in the observing band. The
dashed circle represents the TAB test area and stars depict positions of
randomly generated sources. Two red circles represent a
3-dB contour of a TAB at the higher frequency $\nu_H$ of the observing band.}
\label{fig:lofar_beam_grid_nyq}
\end{figure}

\subsection{Simple Example}
\label{sec:simple_example}
In Fig. \ref{fig:must_ratio_maps} we graphically illustrate the methodology
of finding a source location using the observed S/N pattern for the MUST array.
We simulated a source at the position $[-2.1^{\circ},-3.8^{\circ}]$ from  the
assumed FoV centre in two overlapping TABs. This is marked with a black star in
Fig. \ref{fig:ratio_maps_must} (top) and in Fig.
\ref{fig:must_ratio_maps}. The source was detected in two TABs and at $\nu_H$
and $\nu_L$ fulfilling both requirements of our restrictive definition of
detection. As we are demonstrating only a simple two TAB detection the
subscripts $(i,\;j)$ are omitted here. The observed S/N pattern yields
$\text{S}_1/\text{S}_2$ = 0.86 and $(\alpha_{O_1}-\alpha_{O_2})$ = - 0.52, for
the observed flux density ratio and the observed spectral index difference
respectively. In this first illustration of the technique we consider first
only modest errors $(\pm1\%)$ on the calculated values of
$\mathfrak{s}_1/\mathfrak{s}_2$ and $(\mathfrak{a}_2 - \mathfrak{a}_1)$ and do
not include any statistical error propagation for ease of presentation.  A
further error discussion is included below while a detailed analysis will be
given in a later paper.\\  

Fig. \ref{fig:must_ratio_maps} (top) shows the repeating values of
$\mathfrak{s}_1/\mathfrak{s}_2 = S_1/S_2 \pm1\%$ from the 2-D flux density
ratio $\mathscr{S}_{1}/\mathscr{S}_{2}$ map. The repeating regions of
$\mathfrak{s}_1/\mathfrak{s}_2$ from  the TAB pairs are indicated by the colour
green and the TAB contours are plotted as well to guide orientation.  Fig.
\ref{fig:must_ratio_maps} (middle) shows, plotted in blue, the repeating values
of $(\mathfrak{a}_2 - \mathfrak{a}_1) = (\alpha_{O_1} - \alpha_{O_2}) \pm1\%$
obtained from the 2-D instrumental spectral index difference $(\mathscr{A}_2 -
\mathscr{A}_1)$ map. If we overlay the top and the middle panel, as in Fig.
\ref{fig:must_ratio_maps} (bottom), we can establish a position where both
these allowed regions, $(\mathfrak{a}_2 - \mathfrak{a}_1)$ and
$\mathfrak{s}_1/\mathfrak{s}_2$, overlap. \emph{Only those values that share
the same coordinates are treated as a possible source location}. In our
example, there is only one common coordinate, for both values of
$(\mathfrak{a}_2 - \mathfrak{a}_1)$ and $\mathfrak{s}_1/\mathfrak{s}_2$, which
lies at $[-1.95^{\circ},-3.9^{\circ}]$, and thus is marked with a red dot in
Fig. \ref{fig:must_ratio_maps} (bottom). By this means we were able to
estimate the source location to within $0.02$ HPBW$_L$ of a TAB from the true
source position marked with a black star in Fig. \ref{fig:must_ratio_maps}
(bottom). For example, for the LOFAR array an accuracy of $0.02$ HPBW$_L$
signifies 0.5 arcminute distance from the true source position, whereas for
MeerKAT the same distance signifies one arcsecond. By increasing the assumed
error on the calculated $\mathfrak{s}_1/\mathfrak{s}_2$ ratios and the
$(\mathfrak{a}_2 - \mathfrak{a}_1)$ differences with a two TAB detection, the
number of estimated positions will increase. For example, a $5\%$ error on both
the ratio and the difference results in four additional estimated locations.
However, as is shown later in the text, even considerable errors still allow a
useful location estimation if a source is detected with several TABs treated
in a pairwise fashion.\\
\subsection{Error considerations}
\label{sec:errors}
In the simplified example just shown we assumed highly significant detections
(i.e. S/N of 100) simply to illustrate the method. These do not reflect likely
S/Ns in actual observations so here we consider that, to qualify as a
detection, a transient source must have S/N $\geqslant8$ and thus a related
flux error of $\delta S \leqslant 12.5\%$. The question of whether the errors
between TABs are correlated or independent is non-trivial, being a combination
of several factors. For example there may be a degree of non-independent noise,
depending on the relative contribution of the sky and receiver temperatures;
these will be quite different in the LOFAR and MeerKAT arrays. We will address
the issue of sources of errors and their treatment in a later paper. For the
present discussion we consider the worst case scenario for all three arrays
where the errors taken to be highly correlated and hence
combine linearly rather than quadratically. However, for comparison, we
also show the results from considering the errors to be independent for the
MeerKAT core (Table \ref{table:appendix}). It has good range of baselines and
is dominated by the receiver noise at the assumed frequency of 1.4 GHz.\\ 

The error $\delta (S_1/S_2)_{\nu}$ on the observed flux density ratio
$(S_1/S_2)$ comes from the S/N of the TAB detections, i.e.:
\begin{equation}
        \delta (S_1/S_2)_{\nu} = \left(\frac{\delta S_1}{S_1} + \frac{\delta
        S_2}{S_2}\right)_{\nu}.
\end{equation}
The observed $(S_1/S_2)$ and the modelled $\mathfrak{s}_1/\mathfrak{s}_2$
ratios, obtained from the 2-D flux density ratio
$\mathscr{S}_{1}/\mathscr{S}_{2}$ map, can then be compared. We are looking for
regions where the modelled $\mathfrak{s}_1/\mathfrak{s}_2$ falls within the
range:
\begin{equation}
        S_1/S_2 \pm \delta (S_1/S_2). 
\end{equation}
For the sources simulated in this study, the cumulative error on the flux ratio
$(S_1/S_2)$ is typically $\pm20\%$.\\

The measured flux densities are is also used via Equation
\ref{eq:spectral_index_mod} to estimate the observed spectral index $\alpha_O$.
The cumulative error $\delta \alpha_O$ is calculated as follows:
\begin{equation}
        \delta \alpha_O = \frac{0.43}{\log
        \left({\frac{\nu_L}{\nu_H}}\right)} \left(\frac{\frac{\delta
        S_{\nu_{L}}}{S_{\nu_{L}}} + \frac{\delta
        S_{\nu_{H}}}{S_{\nu_{H}}}}{\frac{S_{\nu_{L}}}{S_{\nu_{H}}}}\right).
\end{equation}
The error then propagates further as we are considering the
observed spectral index difference from two TABs (Equation \ref{eq:si3}):
\begin{equation}
        \delta (\alpha_{O_2} - \alpha_{O_1}) = \delta \alpha_{O_2} + \delta
        \alpha_{O_1}. 
\end{equation}
The calculated $(\alpha_{O_2} - \alpha_{O_1})$ and modelled $(\mathfrak{a}_2 -
\mathfrak{a}_1)$, from the 2-D the spectral index difference $(\mathscr{A}_2 -
\mathscr{A}_1)$ map, are then compared, we are looking for regions where the
modelled $(\mathfrak{a}_2 - \mathfrak{a}_1)$ falls within the range:
\begin{equation}
        (\alpha_{O_2} - \alpha_{O_1})
        \;\pm\; \delta (\alpha_{O_2} - \alpha_{O_1}). 
\end{equation}
The cumulative error on the spectral index difference $\delta (\alpha_{O_2} -
\alpha_{O_1})$ is typically large in our simulations. It is at least $60\%$ and
for the present purpose we have limited the error to $100\%$. It is important
to note that these error refer only to the simulated sources presented in this
work and are not generic to the method. These errors can be  reduced with large
fractional bandwidths but that can make a detection at two frequencies
difficult.
\subsection{Simulation Parameters}
To explore the capabilities of our method, we tested three different spatial
sampling methods to determine how important the TAB separation is for the
accuracy of determining a source location. We first consider the case where the
TABs undersampled the FoV. This means that TABs touch at the $\text{HPBW}_L$
contour at the lowest observing frequency $\nu_L$. In the second case, the TABs
Nyquist sample the FoV i.e. the separation between the phase centres of TABs is equal
to $\text{HPBW}_L/2$ at the lower edge of the observed frequency band. Of
course Nyquist sampling at $\nu_L$ means undersampling at $\nu_H$. In the third
case, the TABs oversample the FoV i.e. the TABs are Nyquist sampled at the
highest frequency ($\text{HPBW}_H/2$). Oversampling is used as a control to
determine if there are any significant benefits from a high surface density of
TABs.\\

Fig. \ref{fig:lofar_beam_grid_nyq} shows an example of the generic test setup
used in the analysis.  We only show the case of Nyquist-sampled TABs as this
layout is similar for all arrays\footnote{The undersampled TAB positions are
also generic for all arrays but are not shown here due to space constrains.
Only the oversampled TABs configuration differ for all arrays as the fractional
bandwidths also differ.}.  Fig. \ref{fig:lofar_beam_grid_nyq} also shows a 3-dB
contour of a TAB at the higher frequency $\nu_H$ in the LOFAR Superterp
observing band to illustrate the undersampling at $\nu_H$. The independent TABs
are simulated to occupy a circle with a diameter equal to
$3\times\text{HPBW}_L$. The centre of the circle is located at the zenith for
each telescope's geographical location and for a specific MJD. A set of 60
strong point sources, all with $\alpha_{I}$ equal to -2, is randomly
distributed in a square of side equal to $3\times\text{HPBW}_L$ encompassing
the TABs. The predicted detection significance at the TAB centre is
taken to be $20\sigma$ for each source at $\nu_L$ and scaled by the spectral
index to obtain the flux density at $\nu_H$. We then calculated the TAB
sensitivities using Equation \ref{eq:srms}.\\

In our simulations we have assumed a perfectly calibrated array where
all the antenna gains are uniform and our beam model is ideal. In a real
telescope, each antenna will have uncalibrated errors in gain and phase. This
inevitably will affect the performance of our method, but for this
proof-of-concept analysis we assume perfect calibration. In Section 5 we identify
several other practical issues which will need to be quantified in a real-life
application of this method. 

\section{Results}
\begin{figure}
\centering
\includegraphics[width=\linewidth]{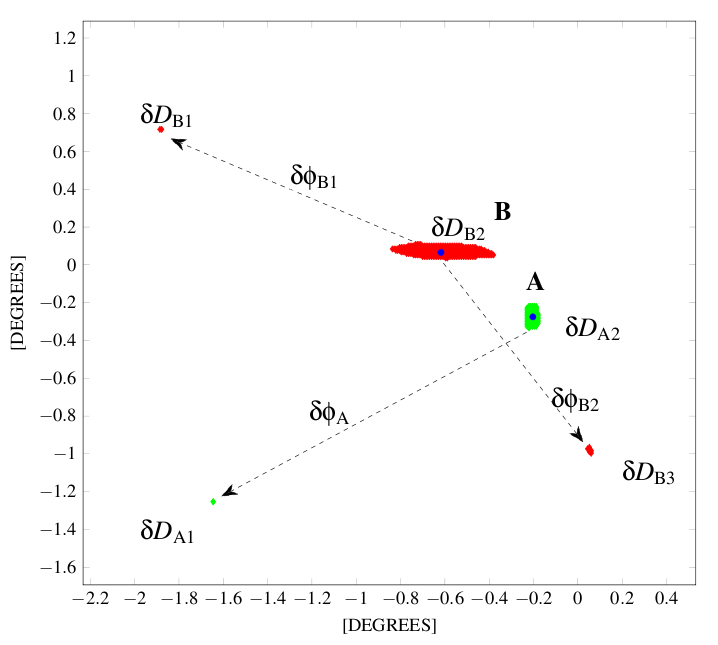}
\caption{Schematic representation of locations \textit{m}, the total area
\textit{D} and the angular distance $\delta\phi$ from the true source position,
marked with blue dot.  The green patches ($m=2$) represent the two estimated
locations for source A.  Red patches ($m=3$) represent the three estimated
locations for source B.  While there may be a significant angular distance
between the estimated locations, the areas to search at these locations may be
quite small.} \label{fig:table_graph} \end{figure}

To present the results of our simulations in a compact form, we divided the
data according to how many locations \textit{m} were estimated per detected
source. Location is used to describe a region where the true
source position might be. For the purpose of illustration, Fig.
\ref{fig:table_graph} is a schematic showing estimated locations for a
hypothetical source A with $m=2$ and a source B with $m=3$. Table
\ref{table:res_2} lists sources for which the number of estimated locations
\textit{m} is between 1 and 5 and also a group containing all detections with
more than 5 estimated locations.  For the number of sources \textit{n} in each
location group we calculated the mean number of TABs ($\overline{B}$) in which
the source was detected to illustrate the benefits of detection with many TABs.
Next we list the minimum ($\delta D_{Min}$), maximum ($\delta D_{Max}$) and
mean ($\overline{D}$) value of the total area \textit{D} covered by the
overlapping values of $\mathfrak{s}_1/\mathfrak{s}_2$ and $(\mathfrak{a}_2 -
\mathfrak{a}_1)$, normalised to the TAB's solid angle\footnote{Here, the solid
angle term is used in relation to a TAB cross section at a HPBW point excluding
sidelobes.} $\Omega_{HPBW_L}$ at $\nu_L$. The total area \textit{D} represents
the area that would need to be sampled in follow-up, or archival, observations
to be sure that the actual position of the source was observed. In our
simulations, the smallest \textit{D} is represented by a single pixel, i.e. the
smallest resolution element of the simulated beam pattern. A single pixel
covers an area of $0.32$ square arcmin for the MUST array, $0.23$ square arcsec
for the LOFAR Superterp and $2.25\times10^{-4}$ square arcsec for
MeerKAT.\\

In Fig. \ref{fig:table_graph} it is clear that the total area $\delta D_B$ for
a source B is considerably larger than the total area $\delta D_A$ for source
A. However, the total area $\delta D_B$ is condensed to mostly a single patch.
Out of the two, the location estimated for source A has lower uncertainty as
there are only two estimated locations and $\overline{D}$ is smaller than for
source B.\\  

In Table  \ref{table:res_2} we also list the minimum ($\delta\phi_{Min}$),
maximum ($\delta\phi_{Max}$) and mean ($\overline{\Delta\phi}$) value of the total
angular distances $\Delta\phi$ between the true and the estimated source
positions, normalised to the HPBW$_L$. We illustrate the above using an example
from column two in Table \ref{table:res_2}, for the MeerKAT array using the
Nyquist sampled method; two patches (\textit{m}) were estimated for three
sources (\textit{n}) with a mean number of TABs ($\overline{B}$) equal to three. The
maximum angular distance $\delta\phi_{max}$ for a detection in this group is
6.3 HPBW$_L$ away from the true position but the mean area $\overline{D}$ to survey
for possible host galaxies is only 0.12 of the $\Omega_{HPBW_L}$ area. This
would suggest a similar scenario to source A (small $\overline{D}$ and large
$\delta\phi_{max}$) in Fig. \ref{fig:table_graph}.\\

We now review the overall statistics of the results starting with the detection
rates.  We then give examples of detections with low and high positional
uncertainty, followed by an estimation of the accuracy of recovery of the true
source position. We finish by presenting the results of the intrinsic spectral
index recovery.
\subsection{Detection Rates}
The number of detected sources that meet our two criteria for detection, are
summarised in Table \ref{table:res_1}. For the LOFAR and MeerKAT arrays with
large fractional bandwidths (23$\%$ and 19$\%$ respectively) the undersampling
of the FoV yields the lowest detection rate of two and three sources out of 60
respectively. This is not unexpected as the first condition for detection is
difficult to meet without overlapping TABs. For the MUST array, with strong
regular sidelobes and small fractional bandwidth of 8$\%$, the undersampling of
the FoV yields a detection of 14 out of 60 random sources. Because of our
restrictive definition of detection and due to the low detection rates for the
LOFAR and the MeerKAT arrays, we will no longer consider the results from
the undersampling method.\\  

Sampling of the FoV with the Nyquist-sampled TABs
increased the detection rates dramatically for all arrays. The biggest change
is observed for the LOFAR array which detected 48 sources. The lowest detection
rate is achieved with the MeerKAT array at only 33 sources out of 60 sources.
As discussed below this is likely due to the high fractional bandwidth, as
illustrated in Fig. \ref{fig:lofar_beam_grid_nyq}. This effect is not
replicated in LOFAR, with similar fractional bandwidth, due to the presence
of strong sidelobes, as illustrated in Fig. \ref{fig:sens_maps}.\\

\begin{figure}
\centering
\includegraphics[width=\linewidth]{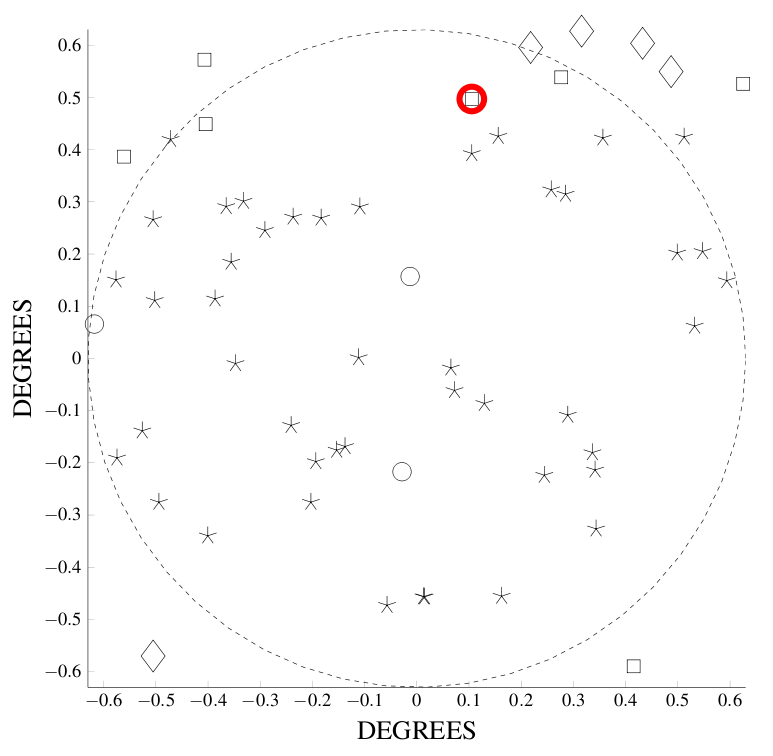}
\caption{Graphical representation of sources detected using different sampling
methods for the LOFAR Superterp TABs. Black circle: sources detected with
undersampling, Nyquist sampling and oversampling. Star: sources detected
with Nyquist sampling and oversampling. Diamond: no detection with any
sampling method. Square: sources detected with oversampling only.}
\label{fig:lofar_det}
\end{figure}

Fig. \ref{fig:lofar_det} shows a representation of sources detected using
different sampling methods for the LOFAR TABs. All undetected sources
($\Diamond$) lay outside of the TAB test area. But some sources that lie
outside of the test area were detected with the oversampling of the FoV.  A
source marked with a red circle in Fig. \ref{fig:lofar_det} and in Fig.
\ref{fig:lofar_beam_grid_nyq}, illustrates the impact of fractional bandwidth
on the detection rates. Fig. \ref{fig:lofar_beam_grid_nyq} shows the Nyquist
sampled FoV with the LOFAR TABs at the lower frequency $\nu_L$ (black circles),
where two contours of TABs at the higher frequency $\nu_H$ in the observing
band are also shown (red circles). The marked source is not detected when the
FoV is Nyquist sampled as the second condition for successful detection is not
satisfied. For simplicity our simulations considered only two
frequencies, at the extremes of the bandwidth. The detection rate could be
improved by considering many sub-bands and in a future paper we plan to discuss
how this would, in turn, improve the location accuracy.

\begin{table*}[p]\centering
\scriptsize
\caption{Comparison of the accuracy of the Nyquist sampled and oversampled
methods for determining the positions and the intrinsic spectral index. Here
\textit{m} is the number of locations estimated per detected source. \textit{n}
is the number of sources detected at the number of positions shown.
$\overline{B}$ is the mean number of TABs per source. Below we list the minimum
($\delta D_{Min}$), maximum ($\delta D_{Max}$) and mean ($\overline{D}$) value
of the total area \textit{D} covered by the overlapping values of
$\mathfrak{s}_1/\mathfrak{s}_2$ and $(\mathfrak{a}_2 - \mathfrak{a}_1)$,
normalised to the $\Omega_{HPBW_L}$ area in deg$^2$. The total area \textit{D}
represents the area that would need to be sampled in follow up, or archival,
observations to be sure that the actual position of the source was observed. We
also list the minimum ($\delta\phi_{Min}$), maximum ($\delta\phi_{Max}$) and
mean ($\overline{\Delta\phi}$) value of the total angular distances
$\Delta\phi$ between the true and the estimated source position, normalised to
the HPBW$_L$.  $\overline{\alpha_I}$ is the mean estimated intrinsic spectral
index of all sources from group \textit{m} where $\sigma$ is the standard
deviation from that mean.}
\label{table:res_2}
\ra{1.3}
\begin{tabular}{@{}lllllllllllllll@{}}\toprule
& &\multicolumn{13}{c}{MUST}\\
\multicolumn{2}{c}{Parameter} &\multicolumn{6}{c}{Nyquist sampling} & \multicolumn{6}{c}{Oversampling}\\
 \cmidrule{3-8} \cmidrule{10-15}
\footnotesize{m}&&1&2&3&4&5&>5& &1&2&3&4&5&>5\\ \midrule
\footnotesize{n}&&33&1&3&1&1&1& &30&4&1&-&1&4\\
\footnotesize{$\overline{B}$}&&5&2&2&3&2&8& &6&3&3&-&2&6\\
\footnotesize{D}\\
&\footnotesize{$\delta D_{Min}$}&0.00045&0.084&0.00023&0.00023&0.043&0.00023& &0.00023&0.00023&0.00023&-&0.00023&0.00023\\
&\footnotesize{$\delta D_{Max}$}&1.4&0.25&1.5&0.062&0.73&0.00023& &0.42&1.6&0.03&-&0.6&0.8\\
&\footnotesize{$\overline{D}$}&0.08&0.17&0.30&0.02&0.41&0.00023& &0.05&0.28&0.01&-&0.20&0.07\\
\footnotesize{$\delta\phi$}\\
&\footnotesize{$\delta\phi_{Min}$}&0.00062&0.0018&0.0038&0.0052&0.0035&0.052& &0.0012&0.0010&0.0018&-&0.0024&0.0017\\
&\footnotesize{$\Delta\phi_{Max}$}&0.7&1.6&1.6&0.2&3.5&1.6& &0.4&1.7&0.8&-&3.1&2.2\\
&\footnotesize{$\overline{\Delta\phi}$}&0.04&0.65&0.25&0.08&1.3&0.72& &0.04&0.22&0.05&-&0.70&0.60\\
\footnotesize{$\alpha_I$}\\
&\footnotesize{$\overline{\alpha_I}$}&-2.0&-2.3&-1.8&-2.0&-1.5&-0.3& &-1.9&-2.1&-2.0&-&-0.9&-0.8\\
&\footnotesize{$\sigma$}&0.2&1.0&0.8&0.6&1.7&2.5& &0.3&0.5&0.4&-&1.8&1.8\\
\bottomrule
& &\multicolumn{13}{c}{LOFAR}\\
\multicolumn{2}{c}{Parameter} &\multicolumn{6}{c}{Nyquist sampling} & \multicolumn{6}{c}{Oversampling}\\
 \cmidrule{3-8} \cmidrule{10-15}
\footnotesize{m}&&1&2&3&4&5&>5& &1&2&3&4&5&>5\\ \midrule
\footnotesize{n}&&7&6&4&3&2&26& &24&6&1&3&2&19\\
\footnotesize{$\overline{B}$}&&4&3&3&3&3&3& &5&4&5&3&3&3\\
\footnotesize{D}\\
&\footnotesize{$\delta D_{Min}$}&0.0004&0.0004&0.0004&0.0004&0.0004&0.0004& &0.0004&0.0004&0.0004&0.0004&0.0004&0.0004\\
&\footnotesize{$\delta D_{Max}$}&0.6&0.6&1.1&0.3&0.7&2.9& &0.4&0.1&0.0004&0.8&0.2&1.6\\
&\footnotesize{$\overline{D}$}&0.09&0.08&0.2&0.05&0.09&0.07& &0.03&0.03&0.0004&0.14&0.03&0.07\\
\footnotesize{$\delta\phi$}\\
&\footnotesize{$\delta\phi_{Min}$}&0.006&0.004&0.007&0.003&0.007&0.002& &0.002&0.004&0.003&0.004&0.004&0.003\\
&\footnotesize{$\delta\phi_{Max}$}&1.3&4.1&3.7&3.8&3.9&8& &0.7&2.9&0.04&3.6&3.8&8.1\\
&\footnotesize{$\overline{\Delta\phi}$}&0.09&0.6&0.5&0.4&0.6&1.6& &0.05&0.4&0.03&0.6&0.6&1.7\\
\footnotesize{$\alpha_I$}\\
&\footnotesize{$\overline{\alpha_I}$}&-2&-1.8&-1.5&-1.8&-1.7&-1.1& &-2&-2&-2&-1.6&-1.8&-1.3\\
&\footnotesize{$\sigma$}&0.12&0.3&0.6&0.3&0.6&1.1& &0.07&0.15&0.09&0.7&0.25&0.8\\
\bottomrule
& &\multicolumn{13}{c}{MeerKAT}\\
\multicolumn{2}{c}{Parameter} &\multicolumn{6}{c}{Nyquist sampling} & \multicolumn{6}{c}{Oversampling}\\
 \cmidrule{3-8} \cmidrule{10-15}
\footnotesize{m}&&1&2&3&4&5&>5& &1&2&3&4&5&>5\\ \midrule
\footnotesize{n}&&25&3&1&2&-&2& &38&2&1&-&1&1\\
\footnotesize{$\overline{B}$}&&4&3&3&3&-&2& &5&3&6&-&2&2\\
\footnotesize{D}\\
&\footnotesize{$\delta D_{Min}$}&0.0004&0.0004&0.0008&0.0004&-&0.0004& &0.0004&0.0015&0.0004&-&0.0008&0.0004\\
&\footnotesize{$\delta D_{Max}$}&2.1&0.6&0.07&2.3&-&2.1& &1.9&2.5&0.0004&-&1.1&1.1\\
&\footnotesize{$\overline{D}$}&0.12&0.12&0.03&0.31&-&0.21& &0.11&0.63&0.0004&-&0.23&0.14\\
\footnotesize{$\delta\phi$}\\
&\footnotesize{$\delta\phi_{Min}$}&0.001&0.002&0.009&0.007&-&0.007& &0.001&0.004&0.007&-&0.008&0.002\\
&\footnotesize{$\delta\phi_{Max}$}&1.3&6.3&9.9&10&-&11& &1.4&6.1&0.04&-&11&11\\
&\footnotesize{$\overline{\Delta\phi}$}&0.08&1.3&0.54&0.66&-&1.1& &0.06&0.36&0.03&-&0.69&0.86\\
\footnotesize{$\alpha_I$}\\
&\footnotesize{$\overline{\alpha_I}$}&-1.9&-1.1&-1.8&-1.5&-&-0.9& &-2&-1.7&-2&-&-1.2&-1.1\\
&\footnotesize{$\sigma$}&0.1&1.5&0.6&0.8&-&1.2& &0.07&0.5&0.05&-&1&1.1\\
\bottomrule
\end{tabular}
\end{table*}

\subsection{Example Detection}
\label{sec:det_example}
\begin{figure*}
        \includegraphics[width=\linewidth/2]{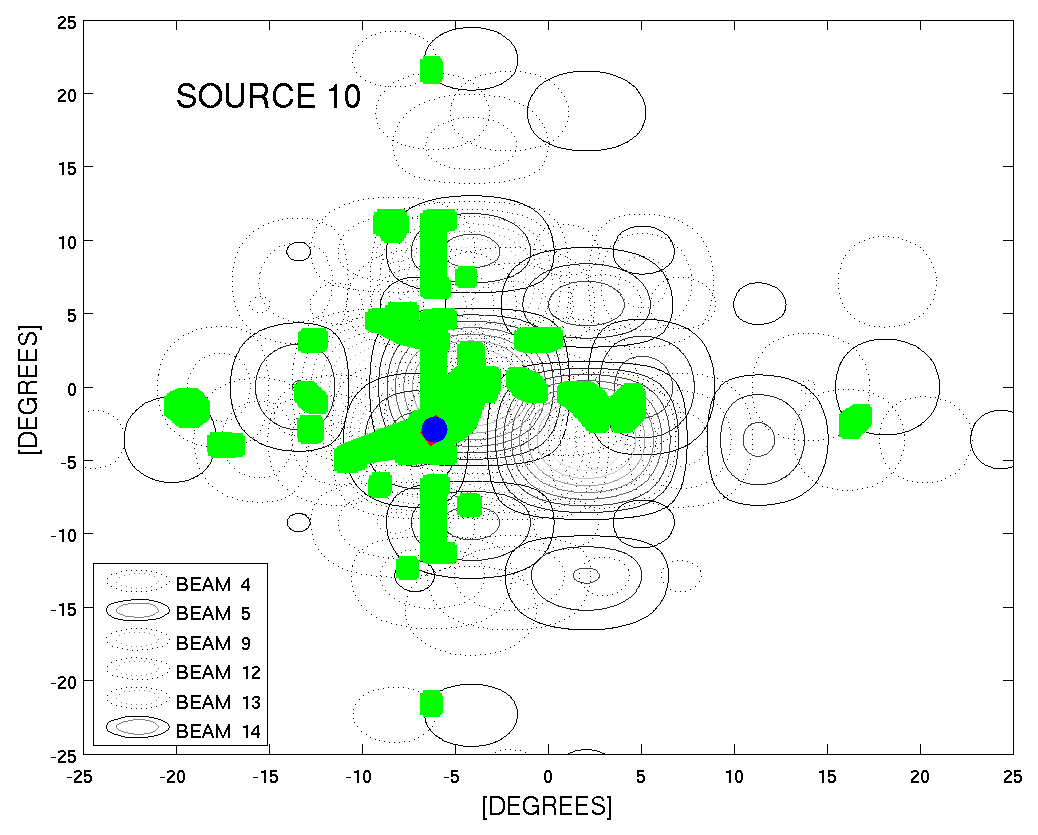}
        \includegraphics[width=\linewidth/2]{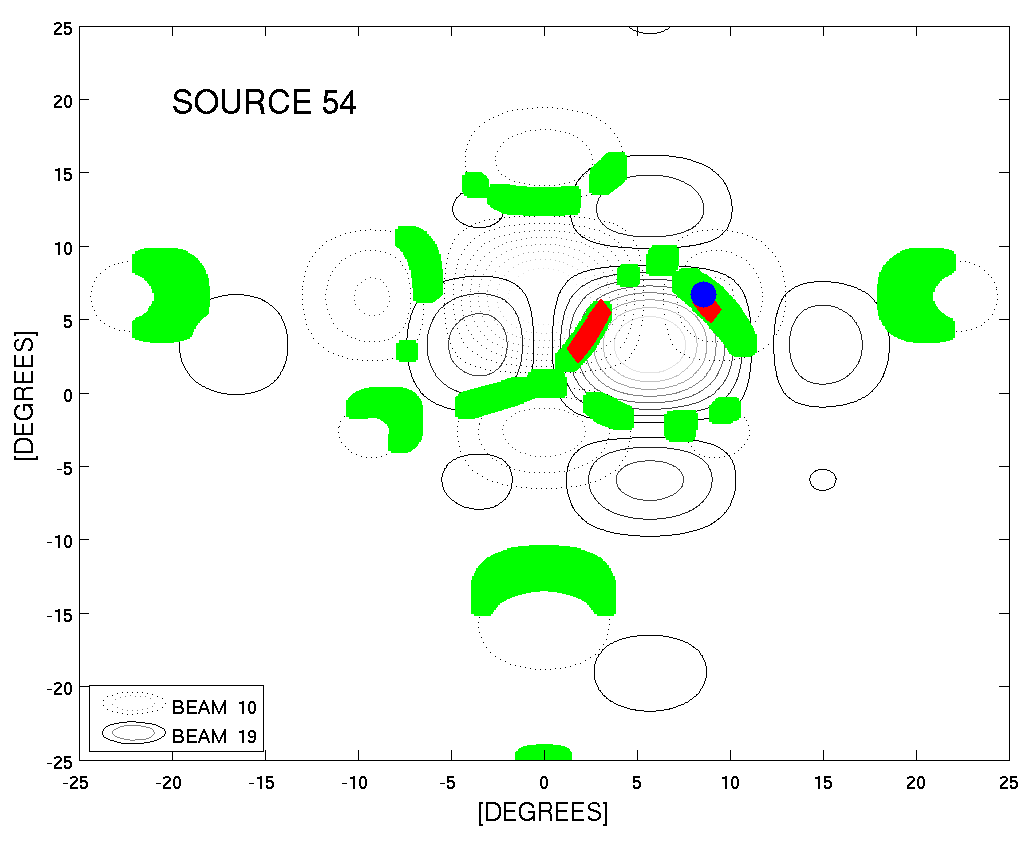}
        \includegraphics[width=\linewidth/2]{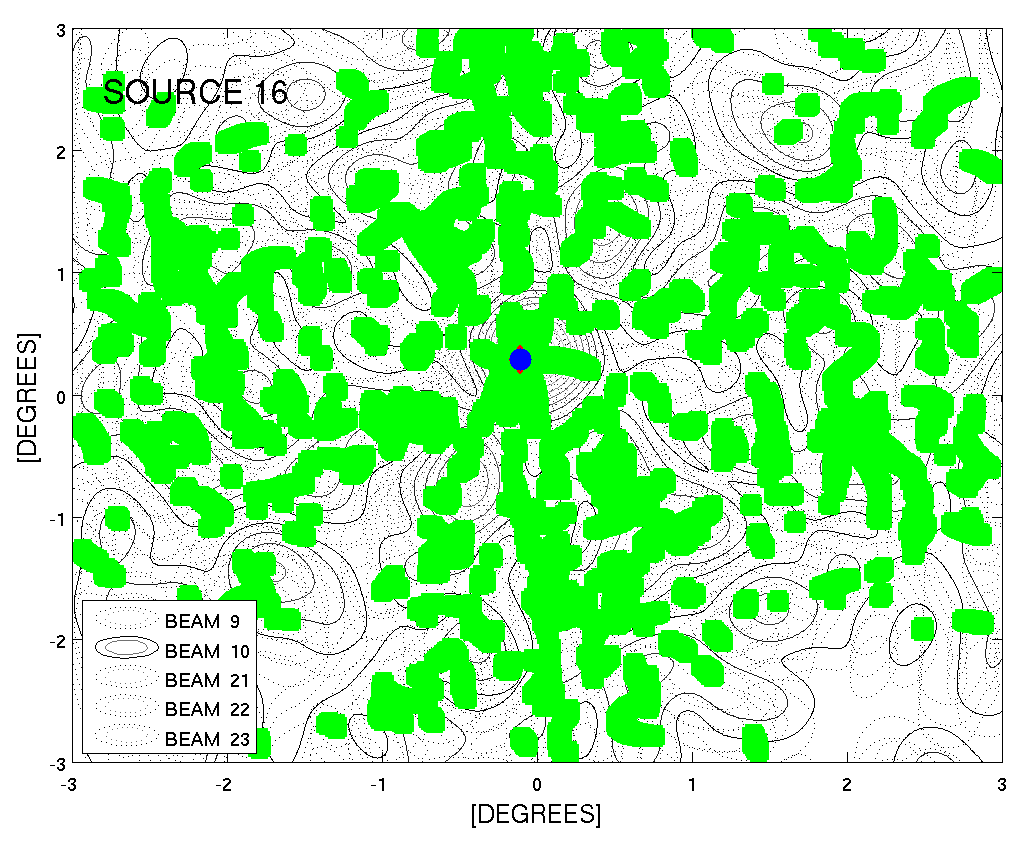}
        \includegraphics[width=\linewidth/2]{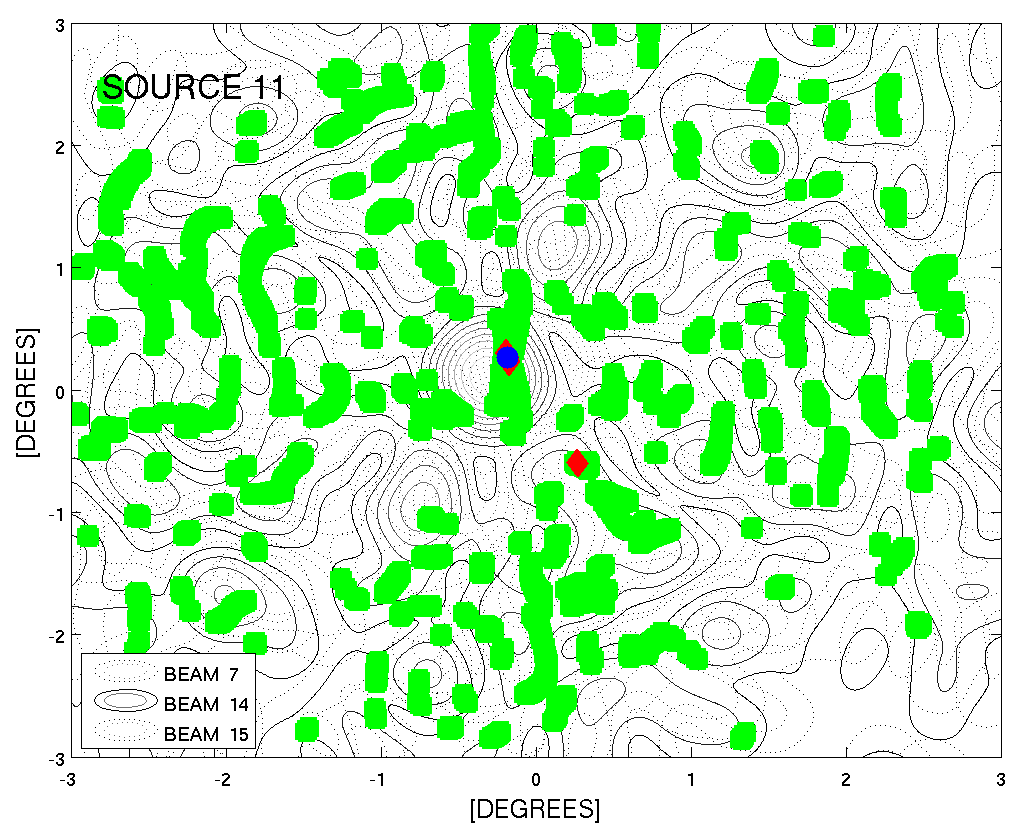}
        \includegraphics[width=\linewidth/2]{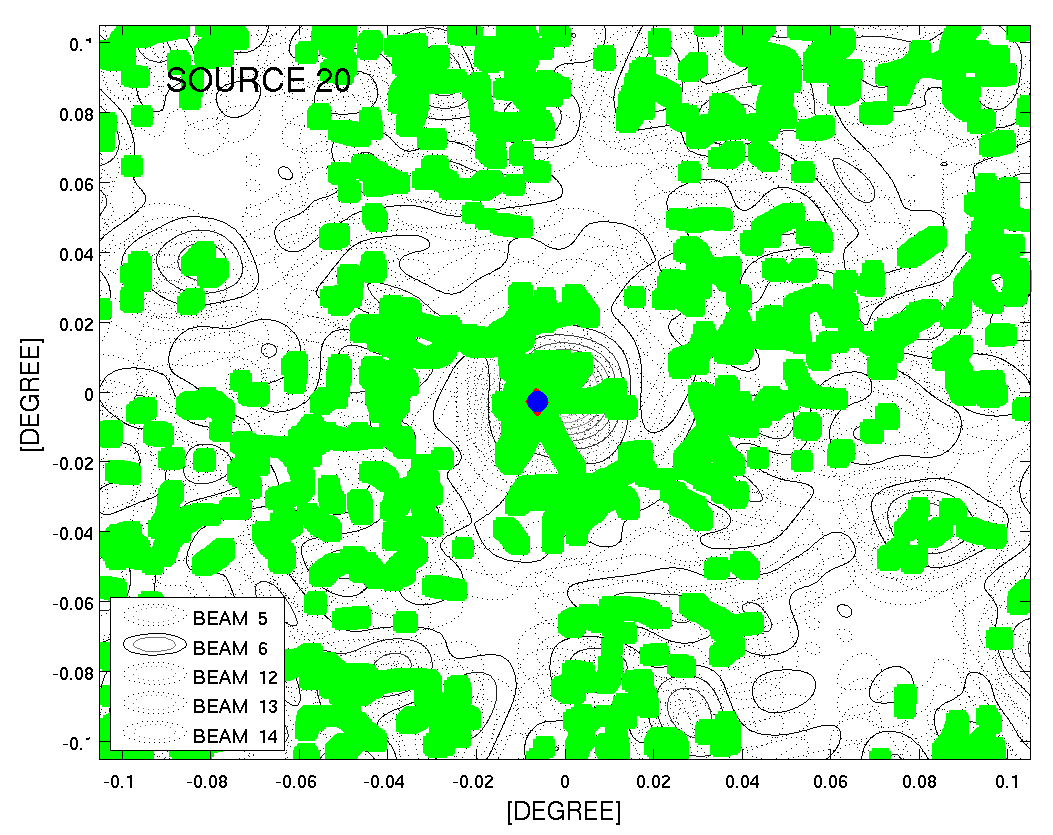}
        \includegraphics[width=\linewidth/2]{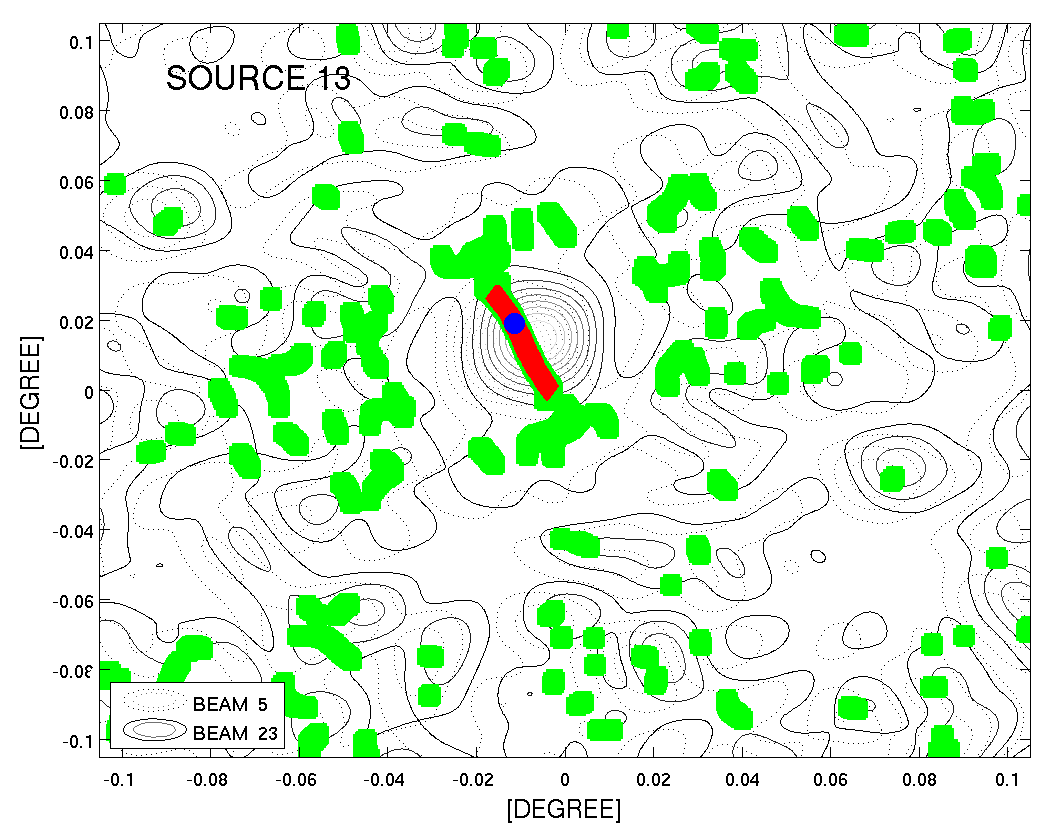}
\caption{Detection examples, discussed in \S\ref{sec:det_example}, with low and
high positional uncertainty (left and right respectively) for the MUST array
(top), the LOFAR array (middle) and MeerKAT array (bottom). (Green squares)
values of $\mathfrak{s}_1/\mathfrak{s}_2$ and $(\mathfrak{a}_2 -
\mathfrak{a}_1)$ from each TAB. The red diamonds regions where
$\mathfrak{s}_1/\mathfrak{s}_2$ and $(\mathfrak{a}_2 - \mathfrak{a}_1)$ from
all TABs overlap. The blue circles the true position of a source. Red regions
in all cases include the true location of the source and so is obscured in the
left panels in particular.}
\label{fig:best_det}
\end{figure*}

In Fig. \ref{fig:best_det} we present examples of the estimated source
positions for representative cases with low (one location only) and high (many
possible locations) positional uncertainty for the MUST array, the LOFAR array and
the MeerKAT array. These also span the range of spatial sampling methods.\\

Each panel in Fig. \ref{fig:best_det} is a simplified version of Fig.
\ref{fig:green_areas} which showed a detection made with five TABs for
illustration only. We use only green squares in Fig. \ref{fig:best_det}
instead to represent overlapping regions from TAB pairs, within the uncertainty
calculated as described in \S\ref{sec:errors}. The red regions show the common
intersection between \emph{all} TAB pairs. The blue circles indicate the true
position of a source, noting that in some cases it obscures the red region. The
contours of $\mathscr{S}$ for each of the TABs where a detection was made are
also plotted.

\subsubsection{Detections with low positional uncertainty}
The top left panel of Fig. \ref{fig:best_det} shows a detection of a source
in six MUST TABs which Nyquist sample the FoV. Our
estimated location of the source is completely consistent with the simulated
position and the total area \textit{D} to scan for a possible counterpart is
only $0.04$ of the $\Omega_{HPBW_L}$. The middle right panel shows a detection
in five LOFAR Superterp TABs which oversample the FoV. The total area \textit{D}
is $0.001$ of the $\Omega_{HPBW_L}$. The bottom right panel shows a detection
in five MeerKAT core TABs which Nyquist sample the FoV, where the total area
\textit{D} is $0.003$ of the $\Omega_{HPBW_L}$. In all these cases the
source was detected in at least five TABs and as a result the number of
repeating values of $(\mathfrak{s}_1/\mathfrak{s}_2)_{(i,\;j)}$ and
$(\mathfrak{a}_2 - \mathfrak{a}_1)_{(i,\;j)}$ from all $(i,\;j)$ pairs of TABs
is substantial (as indicated by the large areas of green). However, this
illustrates why a detection in several TABs helps to accurately pin down its
position as we are looking for values of
$(\mathfrak{s}_1/\mathfrak{s}_2)_{(i,\;j)}$ and $(\mathfrak{a}_2 -
\mathfrak{a}_1)_{(i,\;j)}$ that are common to \textit{all} TABs. 

\subsubsection{Detections with high positional uncertainty}
The top right panel of Fig. \ref{fig:best_det} shows a detection of a source
in two MUST TABs which oversample the FoV. 'Source 54' is detected in a
sidelobe of TAB 10 and the main beam of TAB 19. Unlike for the detections with
low positional uncertainty, only two TABs are contributing
$(\mathfrak{s}_1/\mathfrak{s}_2)$ and $(\mathfrak{a}_2 - \mathfrak{a}_1)$
values. Thus, two possible locations are produced one of which is the real
source position. However, the maximum area $\delta D_{Max}$ to scan for a
counterpart is still only $0.01$ of the $\Omega_{HPBW_L}$.  The middle left panel
shows detection in three LOFAR TABs that Nyquist sample the FoV. Due to the
complicated LOFAR beam pattern the values of
$(\mathfrak{s}_1/\mathfrak{s}_2)_{(i,\;j)}$ and $(\mathfrak{a}_2 -
\mathfrak{a}_1)_{(i,\;j)}$ can be replicated as a result of low level
variations in the beam pattern. However, from that example it is clear that
even a three TAB detection can yield useful results as the maximum area $\delta
D_{Max}$ to scan for a counterpart is only $0.002$ of the $\Omega_{HPBW_L}$. The
bottom left panel shows a detection in two MeerKAT TABs which oversample the
FoV. A two TAB only detection often produces large \textit{D}. This is a common
trend for all arrays. The area \textit{D} for this example is equal to
$1/2\Omega_{HPBW_L}$, but still it offers clues to the true source location and
excludes a detection via a sidelobe.\\

\subsection{Positional Accuracy Estimation}
The results presented in Table \ref{table:res_2} are summarised in Fig.
\ref{fig:nyquist_comp}. In the top panels we show the mean total area $\overline{D}$
covered by the estimated positions normalised to the TAB's solid angle area
$\Omega_{HPBW_L}$, where the error corresponds to one standard deviation from
that mean for each \textit{m}. Points without error bars indicate a single
detection ($n=1$) for that \textit{m}. The middle panels show the mean angular
distance $\overline{\Delta\phi}$ between the estimated and the true source
positions. The accuracy of our method can be best appreciated
when these two sets of panels are examined together. For the MUST and MeerKAT
arrays the positions of the majority of sources, 33 and 25 using Nyquist
sampling and 30 and 38 using the oversampling method respectively, were
estimated with a single position (\textit{m} = 1). For LOFAR array, only 7
sources were estimated with a single position using Nyquist sampling. For 
LOFAR, the oversampling method provided higher accuracy, resulting in detection
of 24 sources with a single position. As illustrated in the middle panels, a single
estimated location on average guarantees a low angular separation $\delta\phi$
from the true source position. For example, for sources in group \textit{m} =
1, the mean total area $\overline{D}$ for the MeerKAT array for both sampling
methods has an area of $10\%$ of $\Omega_{HPBW_L}$ and the mean angular distance
$\overline{\Delta\phi}$ is $8\%$ and $6\%$ of HPBW$_L$ for the Nyquist sampling and
oversampling methods respectively. Looking at the other end of the scale, the
mean total area $\overline{D}$ for the sources
in group \textit{m}$\;>\;$5 is comparable to $\overline{D}$ for sources in group
\textit{m} = 1 for both sampling methods. However, the mean angular distance
$\overline{\Delta\phi}$ of $66\%$ and the mean total area $\overline{D}$ of $86\%$ of
$\Omega_{HPBW_L}$ suggests that the estimated locations include sidelobes as
well.\\ 

The relative results of our method for finding a source location in a TAB,
relative to the HPBW$_L$, favoured the MeerKAT core configuration. This is not
unexpected since the MUST array is a small test array and the complicated beam
pattern of the LOFAR Superterp would require a further development of this
method to get the best out of the data\footnote{An additional metric would need
to be developed to limit the number of estimated source locations and to better
constrain the true position.}. The oversampling of the FoV for the MeerKAT
array gives the best results in terms of small mean total area $\overline{D}$ and
the mean angular distance $\overline{\Delta\phi}$. All 43 sources
detected in oversampled TABs have $\overline{D}$ less than $0.6$ of a HPBW$_L$ area.
For sources in group $m=1$, the locations of 13 sources were estimated within a
single pixel. The mean angular distance for single-pixel sources is 0.3
arcsecond (0.006 of the MeerKAT HPBW$_L$) from the true source position.

\subsubsection{Positional uncertainty for the MeerKAT array when the errors are
independent}
So far all simulations have assumed that the errors between TABs are highly
correlated. In Table \ref{table:appendix} we show for comparison the results
for the MeerKAT array if we consider the errors on
$(\mathfrak{s}_1/\mathfrak{s}_2)_{(i,\;j)}$ and $(\mathfrak{a}_2 -
\mathfrak{a}_1)_{(i,\;j)}$ to be independent. Using again an example for the
Nyquist sampled method the $m=2$ group now contains only one source when the
errors are independent since they are smaller when added quadratically. The
maximum angular distance $\delta\phi_{max}$ for a detection in this group is
1.2 HPBW$_L$ away from the true position. This is a substantial decrease from
$\delta\phi_{max} = 6.3$ (Table \ref{table:res_1}).  In addition, the lower
errors increased the number of source with a single estimated position. For
example for the Nyquist sampled method the number of detections increased from
25 to 29 sources.

\subsection{Intrinsic Spectral Index Recovery}
\begin{figure*}
	\includegraphics[width=\linewidth/2]{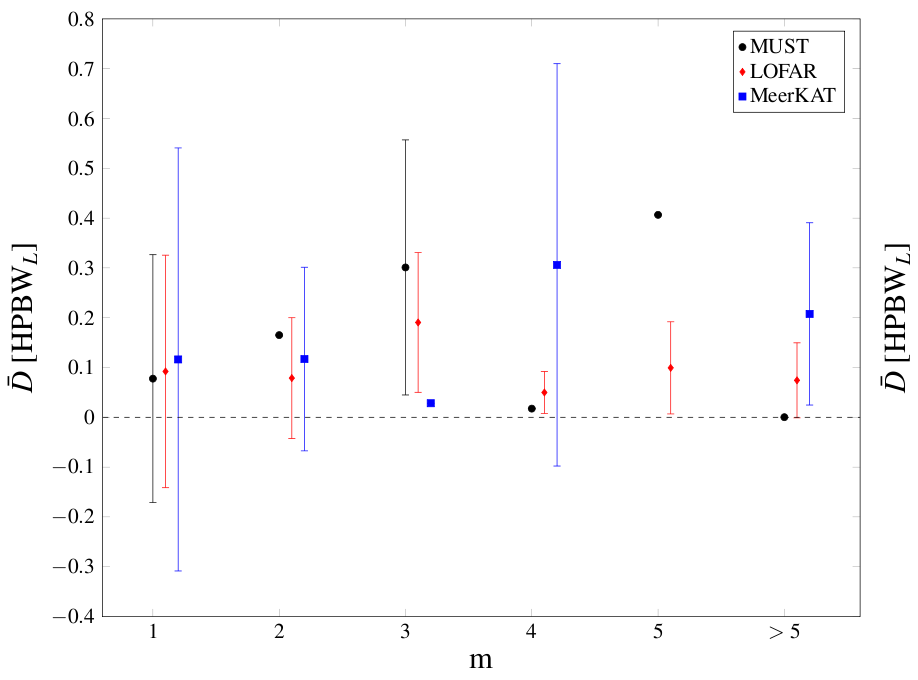}
	\includegraphics[width=\linewidth/2]{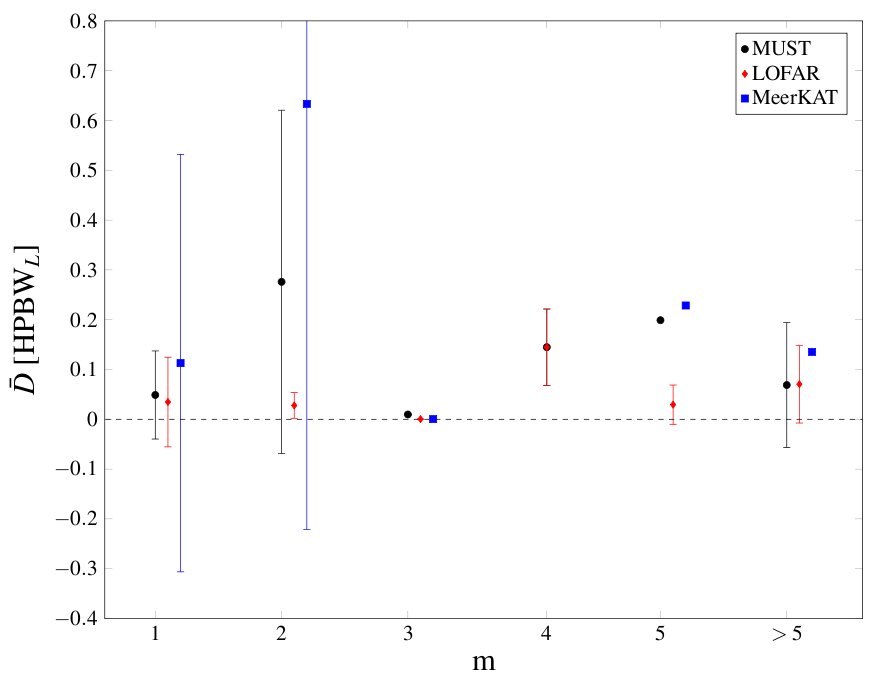}
	\includegraphics[width=\linewidth/2]{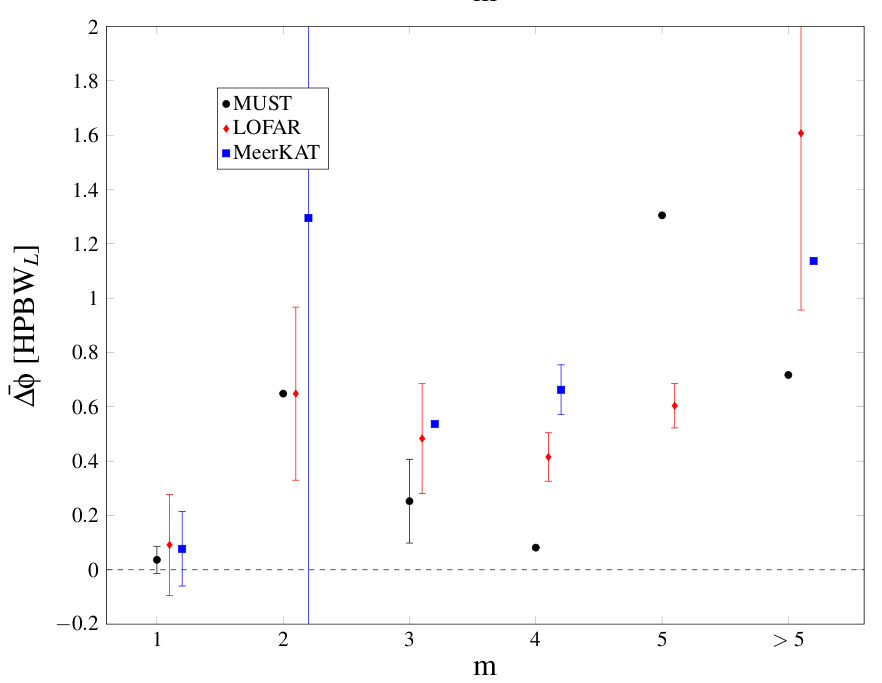}
	\includegraphics[width=\linewidth/2]{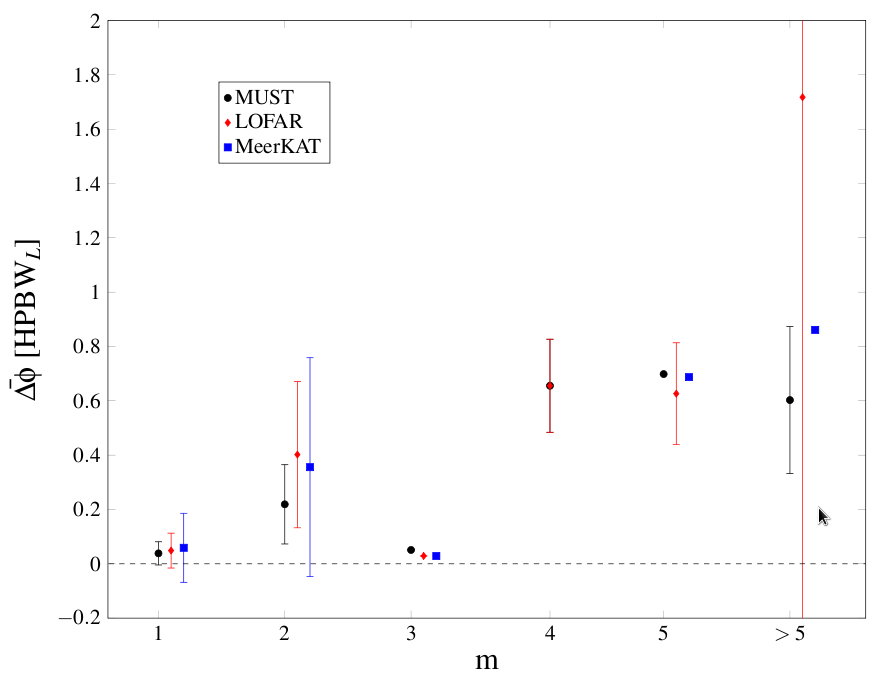}
	\includegraphics[width=\linewidth/2]{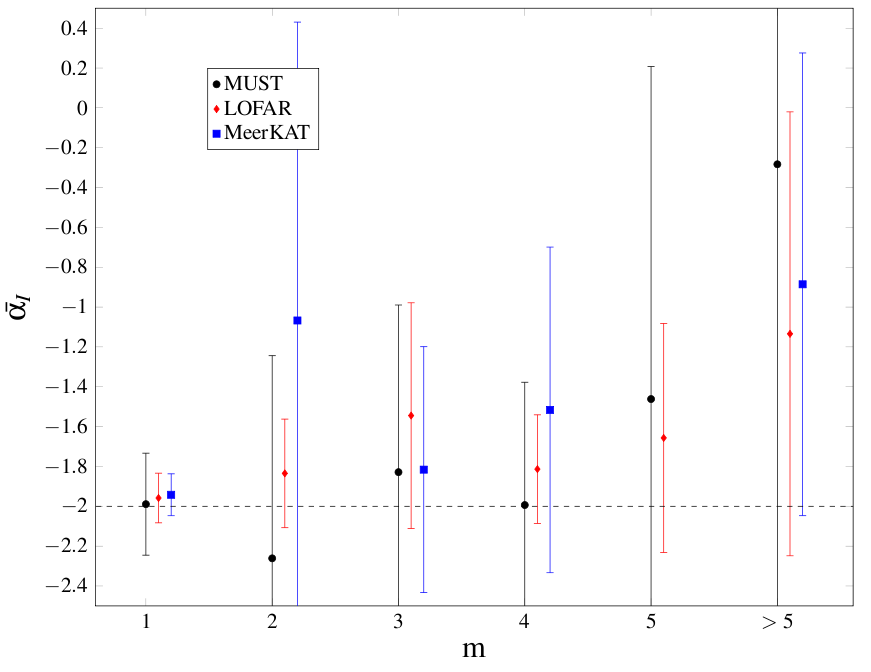}
	\includegraphics[width=\linewidth/2]{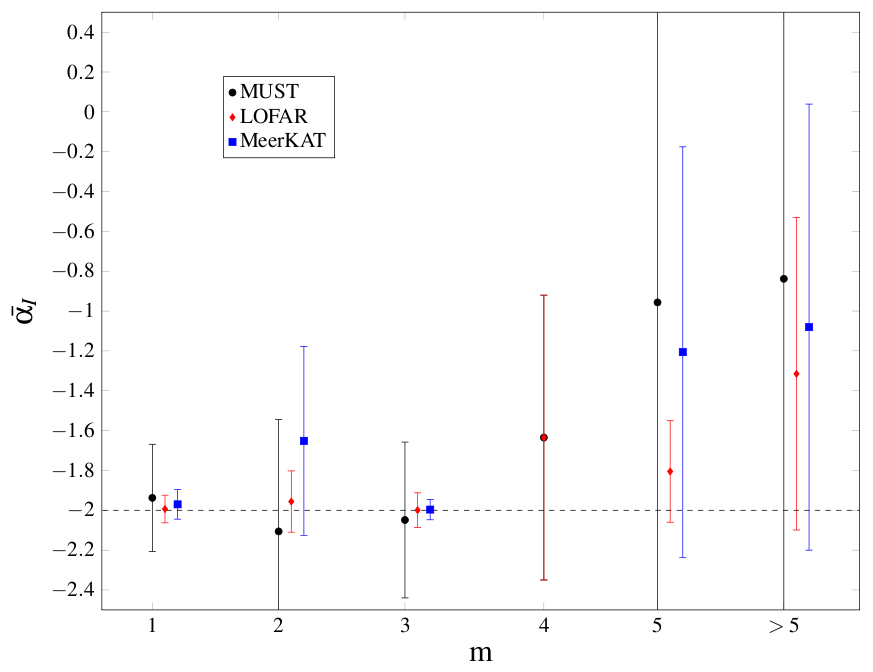}
\caption{Summary of Table \ref{table:res_2} for the Nyquist sampled (left) and
oversampled (right) method. (top) The mean total area $\overline{D}$ covered by the
estimated positions normalised to HPBW area in deg$^2$, where the error
corresponds to one standard deviation from that mean for each \textit{m}.
(middle) The mean angular distance $\overline{\Delta\phi}$ between the estimated and
the true source position.  (bottom) The mean estimated intrinsic spectral index
$\overline{\alpha_{I}}$ for combined sources in each \textit{m}.}
\label{fig:nyquist_comp}
\end{figure*}

A by-product of an accurate position estimation is recovery of the intrinsic
spectral index $\alpha_I$. The last rows in Table \ref{table:res_2} list the
mean estimated intrinsic spectral index $\overline{\alpha_I}$ of a source and the
standard deviation $\sigma$ from that mean from all estimated locations. Fig.
\ref{fig:nyquist_comp} (bottom) shows the mean estimated intrinsic spectral
index $\overline{\alpha_{I}}$ for combined sources in each \textit{m}. For all
arrays where only one small mean area $\overline{D}$ and small $\overline{\Delta\phi}$
were estimated, the recovered $\alpha_{I}$ is close to its assigned value of
-2. Not unexpectedly, there is a clear correlation between the small angular
offset $\delta\phi$ from the true source position and the accuracy of estimated
$\alpha_I$. The more accurate the estimated position is the smaller the
$\alpha_I$ error.

\begin{table}
\caption{Detection rates from a sample of 60 sources.}
\label{table:res_1}
\centering
\begin{tabular}{l c c c}
\hline\hline
Array & Undersampling & Nyquist & Oversampling \\
\hline
   MUST & 14 & 40 & 40 \\
   LOFAR & 3 & 48 & 55 \\
   MeerKAT & 2 & 33 & 43 \\
\hline
\end{tabular}
\end{table}

\section{Discussion}
In summary, we can broadly distinguish three basic levels in a hierarchy of
positional accuracy within a TAB that can be achieved with the method presented
here:
\begin{enumerate}[noitemsep,nolistsep,label=(\roman*)]
\item $\sim 1\times$HPBW accuracy;
\item $\sim 0.1\times$HPBW accuracy;
\item $\sim 0.01\times$HPBW accuracy or better.
\end{enumerate}
For sources that fall into category (i) a detection can be constrained to a
single or small group of TABs. The accuracies of (i) and (ii) may however be
more than sufficient if the observations are conducted in parallel with
follow-up telescopes with higher positional accuracy or in combination with a
transient buffer. On the other hand the angular position of accuracy of (iii),
depending on the maximum baseline, may already be sufficient to identify a host
galaxy, or other related object without reference to other simultaneous
observations.\\

It is clear that a high fractional bandwidth in the case of Nyquist sampling
can prevent one or both of our conditions for detection to be realised. When
the TABs Nyquist sample the FoV, the HPBW contours at $\nu_H$ are spaced
further apart, as illustrated in Fig. \ref{fig:lofar_beam_grid_nyq}, making
our condition (b) difficult to meet. For such arrays oversampling of the FoV
will clearly give higher detection rates. A beam pattern with high number of
low level fluctuations, like LOFAR, can contribute to a higher detection rate,
via detection in sidelobes, but also to a higher number of "false positions".
On the other hand, a clean beam pattern, as produced by MeerKAT, results in
lower detection rates but good location accuracy.  For MeerKAT, the mean
$\overline{\Delta\phi}$ for 38 out of 60 simulated sources, detected in the
oversampled TABs, is less than $6\%$ of the HPBW$_L$, or less than 3 arcsecond
from the true position.  Thus, we were able to accurately estimate the source
location for the majority of sources with a single simulated FRB observation.\\

While the oversampling of the FoV yields superior results for both the LOFAR
and MeerKAT arrays there is a limit on how many TABs can be synthesised with a
given back end. Some coverage of the FoV may therefore have to be sacrificed if
a high location accuracy is desired and this has a trade-off in the survey
speed. For example, say that a FoV is undersampled with \textit{N} TABs, with
\textit{N} being the maximum number of TABs that can be synthesised. If we
survey the same FoV with the Nyquist sampled TABs, the survey would take
approximately 3.5 times longer. If we were to tile the FoV with the oversampled
TABs, the survey would take approximately 4 times longer, for an array with
$10\%$ fractional bandwidth. For an array with $20\%$ fractional bandwidth it
would take 5 times longer.\\

For a real-time transient detection and localisation, the spatial
information measured with sub-second resolution with a correlator
interferometer requires high data rates. The current and the next generation
radio telescopes typically provide both the beamforming mode and the
correlation mode of observation. In that sense, the method we have described
can provide high time resolution and highly useful, sometimes excellent,
positional accuracy without increasing the computational burden of creating an
image. Additionally, there is no significant delay between the detection and
localisation. It is however still of high value to have a transient buffer
available on all dishes in an array, because it will allow for greater
sensitivity by making use of dishes outside of the core. In addition, the
longer baselines will allow even further improvement determination of the
source location. Being able to trigger and store the raw data from the dishes,
and then use the position determined by our methods we can go back and form a
higher sensitivity beamformed data set. This will consist of the raw complex
data which can then be used for coherent dedispersion, in
non-scattering-limited situations, to obtain the true width of the pulse. The
raw complex data can also be used so that the full polarisation calibration,
necessary to achieve the scientific goals mentioned in the introduction, can be
achieved.\\


We stress that this paper describes only a proof-of-concept of the new method.
There are practical issues still to be tackled before the method is ready for
real world observations. In particular a thorough treatment of errors involving
the effect of:
\begin{itemize}[noitemsep,nolistsep]
\item the idealised model of the beam patterns; 
\item the shapes of the receiver bandpasses;
\item the noise contributions of the receivers and sky;
\end{itemize}
will have to be addressed in future work. We also plan to investigate how other
available scientific data, like the DM or polarisation, can be utilised.
Furthermore, our simulations assumed a strict two frequency regime i.e. $\nu_L$
and $\nu_H$, but in practise the bandwidth can be split into many bands
increasing the detection rates for high S/N transients and positively
contributing to the accuracy of the estimated location. Another area of
investigation would be to see whether simultaneously considering many frequency
channels across the band is better than applying a known beam model before
generating values at a smaller number of frequency channels. These steps, we
believe, would also increase the usability of this method for arrays with
complicated beam patterns, like the LOFAR Superterp. The implications of radio
frequency interference (RFI) on the calculation of source flux density will be
addressed in the next paper. However, it is unlikely that RFI will be spatially
distinct compared to real sources so this method may provide further capability
for discriminating against RFI.
\section{Conclusions}
We have presented a proof-of-concept analysis of a new non-imaging method for
detecting and locating radio transient sources. It utilises the additional
spectral and comparative spatial information from multiple TABs formed by an
adding interferometer array to estimate a transient source location in almost
real time. We have shown that this method can work in variety of
interferometers but, not surprisingly, the method is most successful for arrays
with good range of baselines, and hence clean array beam patterns. In this case
transients with high S/N can be localised to small fractions of a HPBW of a TAB
sufficient, in the case of MeerKAT, to localise a source to arcsecond accuracy.
Even less certain positions can still be very useful if there is a parallel
transient search in the same field at a different wavelength. In a future paper
we will address a range of practical issues which will need to be taken into
account in an operational implementation of this scheme.\\

We thank Jayanta Roy for useful discussions and for the help in improving the
clarity of this paper. We would like to thank the referee for providing
comments which serve to improve and clarify the paper. We want to extend our
appreciation for taking the time to comment on our manuscript to the editor. 
\begin{table*}[p]\centering
\scriptsize
\caption{Comparison of the accuracy of the Nyquist sampled and oversampled
methods for determining the positions and the intrinsic spectral index for the
MeerKAT core for the case when the errors are independent. Here \textit{m} is
the number of locations estimated per detected source. \textit{n} is the number
of sources detected at the number of positions shown. $\overline{B}$ is the
mean number of TABs per source. Below we list the minimum ($\delta D_{Min}$),
maximum ($\delta D_{Max}$) and mean ($\overline{D}$) value of the total area
\textit{D} covered by the overlapping values of $\mathfrak{s}_1/\mathfrak{s}_2$
and $(\mathfrak{a}_2 - \mathfrak{a}_1)$, normalised to the $\Omega_{HPBW_L}$
area in deg$^2$. The total area \textit{D} represents the area that would need
to be sampled in follow up, or archival, observations to be sure that the
actual position of the source was observed. We also list the minimum
($\delta\phi_{Min}$), maximum ($\delta\phi_{Max}$) and mean
($\overline{\Delta\phi}$) value of the total angular distances $\Delta\phi$
between the true and the estimated source position, normalised to the HPBW$_L$.
$\overline{\alpha_I}$ is the mean estimated intrinsic spectral index of all
sources from group \textit{m} where $\sigma$ is the standard deviation from
that mean.}
\label{table:appendix}
\ra{1.3}
\begin{tabular}{@{}lllllllllllllll@{}}\toprule
& &\multicolumn{13}{c}{MeerKAT}\\
\multicolumn{2}{c}{Parameter} &\multicolumn{6}{c}{Nyquist sampling} & \multicolumn{6}{c}{Oversampling}\\
 \cmidrule{3-8} \cmidrule{10-15}
\footnotesize{m}&&1&2&3&4&5&>5& &1&2&3&4&5&>5\\ \midrule
\footnotesize{n}&&29&1&1&-&-&2& &39&2&-&-&1&1\\
\footnotesize{$\overline{B}$}&&4&2&2&-&-&2& &5&4&-&-&2&2\\
\footnotesize{D}\\
&\footnotesize{$\delta D_{Min}$}&0.0004&0.0008&0.006&-&-&0.0004& &0.0004&0.0004&-&-&0.0008&0.0004\\
&\footnotesize{$\delta D_{Max}$}&2&0.5&2.1&-&-&1.8& &1.9&2.3&-&-&0.7&0.8\\
&\footnotesize{$\overline{D}$}&0.09&0.27&0.70&-&-&0.19& &0.11&0.58&-&-&0.16&0.13\\
\footnotesize{$\delta\phi$}\\
&\footnotesize{$\delta\phi_{Min}$}&0.001&0.002&0.008&-&-&0.007& &0.001&0.006&-&-&0.008&0.002\\
&\footnotesize{$\delta\phi_{Max}$}&1.3&1.2&6.2&-&-&11& &1.4&6.1&-&-&11&11\\
&\footnotesize{$\overline{\Delta\phi}$}&0.06&0.59&0.72&-&-&0.85& &0.05&0.30&-&-&0.68&0.88\\
\footnotesize{$\alpha_I$}\\
&\footnotesize{$\overline{\alpha_I}$}&-2.1&-1.6&-1.3&-&-&-1.1& &-2.1&-1.7&-&-&-1.2&-1.1\\
&\footnotesize{$\sigma$}&0.1&0.8&0.8&-&-&1.1& &0.1&0.4&-&-&1.1&1.1\\
\bottomrule
\end{tabular}
\end{table*}


\bibliographystyle{mnras}

\begin{thebibliography}{}

\bibitem[\protect\citeauthoryear{{Bannister} \& {Cornwell}}{{Bannister} \&
  {Cornwell}}{2011}]{bc11}
{Bannister} K.~W.,  {Cornwell} T.~J., 2011, \apjs, 196, 16

\bibitem[\protect\citeauthoryear{Bannister et~al.}{Bannister
  et~al.}{2012}]{bmg+12}
Bannister K.~W., Murphy T., Gaensler B.~M.,  Reynolds J.~E., 2012, The
  Astrophysical Journal, 757, 38

\bibitem[\protect\citeauthoryear{{Booth} et~al.}{{Booth} et~al.}{2009}]{bdjf09}
{Booth} R.~S., {de Blok} W.~J.~G., {Jonas} J.~L.,  {Fanaroff} B., 2009, ArXiv
  e-prints

\bibitem[\protect\citeauthoryear{{Burke-Spolaor} et~al.}{{Burke-Spolaor}
  et~al.}{2011}]{bsb11}
{Burke-Spolaor} S., {Bailes} M., {Ekers} R., {Macquart} J.-P.,  {Crawford} F.,
  III, 2011, \apj, 727, 18

\bibitem[\protect\citeauthoryear{{Burke-Spolaor} \&
  {Bannister}}{{Burke-Spolaor} \& {Bannister}}{2014}]{bsb14}
{Burke-Spolaor} S.,  {Bannister} K.~W., 2014, \apj, 792, 19

\bibitem[\protect\citeauthoryear{Cen \& Ostriker}{Cen \& Ostriker}{1999}]{co99}
Cen R.,  Ostriker J.~P., 1999, ApJ, 514, 1

\bibitem[\protect\citeauthoryear{{Chippendale} et~al.}{{Chippendale}
  et~al.}{2010}]{cor+10}
{Chippendale} A.~P., {O'Sullivan} J., {Reynolds} J., {Gough} R., {Hayman} D.,
  {Hay} S., 2010, in Phased Array Systems and Technology (ARRAY), 2010 IEEE
  International Symposium on, pp.648-652, 12-15 Oct. 2010, p. 648

\bibitem[\protect\citeauthoryear{{Cordes} et~al.}{{Cordes}
  et~al.}{2006}]{cfl+06}
{Cordes} J.~M. et~al., 2006, ApJ, 637, 446

\bibitem[\protect\citeauthoryear{{Cordes} \& {Lazio}}{{Cordes} \&
  {Lazio}}{2002}]{col02}
{Cordes} J.~M.,  {Lazio} T.~J.~W., 2002, ArXiv Astrophysics e-prints

\bibitem[\protect\citeauthoryear{{Deng} \& {Zhang}}{{Deng} \&
  {Zhang}}{2014}]{dz14}
{Deng} W.,  {Zhang} B., 2014, \apjl, 783, L35

\bibitem[\protect\citeauthoryear{{Gao}, {Li}, \& {Zhang}}{{Gao}
  et~al.}{2014}]{glz14}
{Gao} H., {Li} Z.,  {Zhang} B., 2014, \apj, 788, 189

\bibitem[\protect\citeauthoryear{{Haslam} et~al.}{{Haslam}
  et~al.}{1982}]{hssw82}
{Haslam} C.~G.~T., {Stoffel} H., {Salter} C.~J.,  {Wilson} W.~E., 1982, A\&AS,
  47, 1

\bibitem[\protect\citeauthoryear{{Hassall}, {Keane}, \& {Fender}}{{Hassall}
  et~al.}{2013}]{hkf13}
{Hassall} T.~E., {Keane} E.~F.,  {Fender} R.~P., 2013, \mnras, 436, 371

\bibitem[\protect\citeauthoryear{{Keane} et~al.}{{Keane} et~al.}{2012}]{kskl12}
{Keane} E.~F., {Stappers} B.~W., {Kramer} M.,  {Lyne} A.~G., 2012, \mnras, 425,
  L71

\bibitem[\protect\citeauthoryear{{Kouwenhoven} \& {Vo{\^u}te}}{{Kouwenhoven} \&
  {Vo{\^u}te}}{2001}]{kv01}
{Kouwenhoven} M.~L.~A.,  {Vo{\^u}te} J.~L.~L., 2001, \aap, 378, 700

\bibitem[\protect\citeauthoryear{{Law} \& {Bower}}{{Law} \&
  {Bower}}{2014}]{lb14}
{Law} C.~J.,  {Bower} G.~C., 2014, in Exascale Radio Astronomy, p. 20303

\bibitem[\protect\citeauthoryear{{Law} et~al.}{{Law} et~al.}{2014}]{lbb+14}
{Law} C.~J. et~al., 2014, ArXiv e-prints

\bibitem[\protect\citeauthoryear{{Law} et~al.}{{Law} et~al.}{2011}]{ljb+11}
{Law} C.~J., {Jones} G., {Backer} D.~C., {Barott} W.~C., {Bower} G.~C.,
  {Gutierrez-Kraybill} C., {Williams} P.~K.~G.,  {Werthimer} D., 2011, \apj,
  742, 12

\bibitem[\protect\citeauthoryear{{Lorimer} et~al.}{{Lorimer}
  et~al.}{2007}]{lbm+07}
{Lorimer} D.~R., {Bailes} M., {McLaughlin} M.~A., {Narkevic} D.~J.,  {Crawford}
  F., 2007, Science, 318, 777

\bibitem[\protect\citeauthoryear{{Lorimer} et~al.}{{Lorimer}
  et~al.}{2006}]{lfl06}
{Lorimer} D.~R. et~al., 2006, \mnras, 372, 777

\bibitem[\protect\citeauthoryear{{Lorimer} et~al.}{{Lorimer}
  et~al.}{2013}]{lkm+13}
{Lorimer} D.~R., {Karastergiou} A., {McLaughlin} M.~A.,  {Johnston} S., 2013,
  ArXiv e-prints

\bibitem[\protect\citeauthoryear{{Macquart} \& {Koay}}{{Macquart} \&
  {Koay}}{2013}]{mk13}
{Macquart} J.-P.,  {Koay} J.~Y., 2013, \apj, 776, 125

\bibitem[\protect\citeauthoryear{{Manchester} et~al.}{{Manchester}
  et~al.}{2006}]{mfl+06}
{Manchester} R.~N., {Fan} G., {Lyne} A.~G., {Kaspi} V.~M.,  {Crawford} F.,
  2006, ApJ, 649, 235

\bibitem[\protect\citeauthoryear{Manchester et~al.}{Manchester
  et~al.}{2001}]{mlc+01}
Manchester R.~N. et~al., 2001, MNRAS, 328, 17

\bibitem[\protect\citeauthoryear{{McQuinn}}{{McQuinn}}{2014}]{mcq14}
{McQuinn} M., 2014, \apjl, 780, L33

\bibitem[\protect\citeauthoryear{{Nan} et~al.}{{Nan} et~al.}{2011}]{nlj+11}
{Nan} R. et~al., 2011, International Journal of Modern Physics D, 20, 989

\bibitem[\protect\citeauthoryear{{Oosterloo}, {Verheijen}, \& {van
  Cappellen}}{{Oosterloo} et~al.}{2010}]{ovv10}
{Oosterloo} T., {Verheijen} M.,  {van Cappellen} W., 2010, in ISKAF2010 Science
  Meeting

\bibitem[\protect\citeauthoryear{Palaniswamy et~al.}{Palaniswamy
  et~al.}{2014}]{pwt+14}
Palaniswamy D., Wayth R.~B., Trott C.~M., McCallum J.~N., Tingay S.~J.,
  Reynolds C., 2014, The Astrophysical Journal, 790, 63

\bibitem[\protect\citeauthoryear{{Petroff} et~al.}{{Petroff}
  et~al.}{2014}]{pbb+14}
{Petroff} E. et~al., 2014, ArXiv e-prints

\bibitem[\protect\citeauthoryear{{Ravi}, {Shannon}, \& {Jameson}}{{Ravi}
  et~al.}{2014}]{rsj14}
{Ravi} V., {Shannon} R.~M.,  {Jameson} A., 2014, ArXiv e-prints

\bibitem[\protect\citeauthoryear{{Reich} \& {Reich}}{{Reich} \&
  {Reich}}{1988}]{rr88b}
{Reich} P.,  {Reich} W., 1988, A\&A, 196, 211

\bibitem[\protect\citeauthoryear{{Spitler} et~al.}{{Spitler}
  et~al.}{2014}]{sch+14}
{Spitler} L.~G. et~al., 2014, ArXiv e-prints

\bibitem[\protect\citeauthoryear{{Thornton} et~al.}{{Thornton}
  et~al.}{2013}]{tsb+13}
{Thornton} D. et~al., 2013, Science, 341, 53

\bibitem[\protect\citeauthoryear{{van Haarlem} et~al.}{{van Haarlem}
  et~al.}{2013}]{vwg+13}
{van Haarlem} M.~P. et~al., 2013, \aap, 556, A2

\bibitem[\protect\citeauthoryear{{Zhou} et~al.}{{Zhou} et~al.}{2014}]{zlw+14}
{Zhou} B., {Li} X., {Wang} T., {Fan} Y.-Z.,  {Wei} D.-M., 2014, \prd, 89,
  107303

\end{thebibliography}



\end{document}